\documentclass[onecolumn,showpacs,preprintnumbers,amsmath,amssymb,nofootinbib,scrartcl]{revtex4}
\input psfig.sty
\usepackage{epsf}
\usepackage{graphicx}  % Include figure files
\usepackage{epstopdf}
\usepackage{dcolumn}   % Align table columns on decimal point
\usepackage{bm}        % bold math
\usepackage{color}

\graphicspath{{./CP/}{./media_z/}}      % Caminho para o diretorio das figuras

             \usepackage{hyperref}
             \hypersetup{colorlinks=true, linkcolor=blue, citecolor=blue, filecolor=blue,
            urlcolor=green,
            pdfauthor={AUTOR DO TRABALHO},
            pdftitle={TITULO DO TRABALHO},
            pdfsubject={TESE ou DISSERTACAO + ANO},
            pdfkeywords={PALAVRAS CHAVES},
            pdfproducer={LaTeX},
            pdfcreator={pdfLaTeX}}
            \usepackage{lastpage}

\newcommand{\be}{\begin{equation}}
\newcommand{\en}{\end{equation}}
\newcommand{\bea}{\begin{eqnarray}}
\newcommand{\ena}{\end{eqnarray}}

\begin{document}

\title{On detecting interactions in the dark sector with $H(z)$ data}

\author{Pedro C. Ferreira$^{1,2,3}${\footnote{pferreira@dfte.ufrn.br}}}

\author{Diego Pav\'{o}n$^2${\footnote{diego.pavon@uab.es}}}

\author{Joel C. Carvalho$^{1, 4}${\footnote{carvalho@dfte.ufrn.br}}}

\affiliation{$^1$Departamento de F\'{\i}sica, Universidade Federal do Rio Grande do Norte,  59072-970 Natal -
Rio Grande do Norte, Brazil}

\affiliation{$^2$ Departamento de F\'{i}sica, Facultad de Ciencias, Universidad Aut\'{o}noma de Barcelona,
08193 Bellaterra (Barcelona), Spain}

\affiliation{$^3$ CAPES Foundation, Ministry of Education of Brazil,
Bras\'{i}lia - DF 70040-020, Brazil}

\affiliation{$^4$Observat\'orio Nacional, 20921-400 Rio de Janeiro
- Rio de Janeiro, Brazil}

\date{\today}

%%%%%%%%%%%%%%%%%%%%%%%%%%%%%%%%%%%%%%%%%%%%%%%%%%%%%%%%%%%%%%%%%%%%%%%%%%%%%%%%
%%%%%%%%%%%%%%%%%%%%%%%%%%%%%%%%%%%%%%%%%%%%%%%%%%%%%%%%%%%%%%%%%%%%%%%%%%%%%%%%
\begin{abstract}
An interesting  approach to the cosmological coincidence problem
is to allow dark matter and dark energy interact with each other
also nongravitationally. We consider two general  {\em
Ans\"{a}tze} for such an interaction and appraise their ability to
address the coincidence problem. We determine the average accuracy
required on the cosmic  expansion rate data to distinguish
interacting cosmological models from the conventional $\Lambda$CDM
scenario. We find that among the planned surveys the Wide Field
Infrared Survey Telescope has the best chance to detect an
interaction, though at a low significance level. To unambiguously
determine the existence of an interaction one must, therefore,
combine the said expansion data with other probes.
\end{abstract}

\pacs{98.80.-k, 95.35.+d, 95.36.+x}

\maketitle

%%%%%%%%%%%%%%%%%%%%%%%%%%%%%%%%%%%%%%%%%%%%%%%%%%%%%%%%%%%%%%%%%%%%%%%%%%%%%%%%
\section{Introduction}
After many years of research very little is known for certain
about the two components of the dark sector: cold dark matter (DM)
and dark energy (DE). The latter, possibly in the form of a
cosmological constant, appears to be the agent driving the current
phase of cosmic accelerated expansion. Thus, some basic questions
immediately arise. For instance, is the cosmological constant a
manifestation of the quantum vacuum? If so, the huge gap  between
the DE density expected from quantum field theory and the observed
one is in urgent need of explanation \cite{Weinberg1989}.
Likewise, why are the densities  of DM and DE of the same order
precisely today? Why are we living in such a special epoch i.e.,
``why now?" This constitutes the so-called ``coincidence problem".
Also, do DM and DE interact gravitationally only or are they
coupled? There is little doubt that  these fundamental questions
must be interrelated whereby shedding light on one of them may
help to explain the others. Following this line of thought we
focus on the why now problem, hoping that if it is better
understood, it will mean a major step in unveiling the nature of
the components of the dark sector. Notice that because of our
current lack of knowledge about the nature of these two components
it would be unwise to discard {\em a priori} an interaction among
them. What is more, they should be coupled unless some so far
unknown-symmetry forbids it. Besides, data obtained from
observations of the dynamics of  galaxy clusters \cite{elcio} and
the integrated Sachs-Wolfe effect \cite{german} hint that indeed
they interact with one another, DE slowly decaying in DM.

To the best of our knowledge,  the first proposal on theoretical
grounds that what we now call dark energy may decay into matter
and radiation was made as early as 1933 by Bronstein
\cite{Bronstein1933}. Later on,  Wetterich \cite{christoff},
Micheletti, Abdalla, and Wang \cite{sandro}, Polyakov
\cite{polyakov}, and Abdalla, Graef, and Wang \cite{elcio-bin},
among others, reinforced the theoretical framework. As we write,
the number of proposals (both from the theoretical as well as from
the phenomenological side) has been greatly augmented -see e.g.
\cite{Overduim1998, Luca2000, Amendola-book}, and references
therein.

Here we shall assume  that the Universe is homogeneous and
isotropic  at large scales, with flat spatial sections, dominated
by nonbaryonic cold matter, by dark energy, and (to a lesser
extent) by baryons (subscripts $m$, $x$, and $b$, respectively).
We neglect radiation, as it results as dynamically unimportant at
the epochs we are interested in.

The overall energy conservation equation reads
%%%%%%%%%%%%%%%%%%%%%%%%%%%%%%%%%%%%%%%%%%%%%%%%%%%%%%%%%%%%%%%%%%%%%%%%%%%%%%%%%%%%%%%%%%%%%%%%
\begin{equation}
    \dot{\rho} + 3H(\rho+p) = 0 ,
\label{eq:conservation1}
\end{equation}
%%%%%%%%%%%%%%%%%%%%%%%%%%%%%%%%%%%%%%%%%%%%%%%%%%%%%%%%%%%%%%%%%%%%%%%%%%%%%%%%%%%%%%%%%%%%%%%%%
where $\rho = \rho_b + \rho_m + \rho_x$, $p$ denotes the total
pressure of the cosmic fluid, and $H$ is the Hubble factor. In the
customary description of the expansion DM and DE evolve
independently, as no interaction other than gravity is considered.
However, there is no apparent reason why they should not interact.
So, in general, they must couple through some  small term $Q$:
%%%%%%%%%%%%%%%%%%%%%%%%%%%%%%%%%%%%%%%%%%%%%%%%%%%%%%%%%%%%%%%%%%%%%%%%%%%%%%%%%%%%%%%%%%%%%%%%%
\begin{equation}
    \dot{\rho}_m + 3H\rho_m = Q \, ,
    \label{eq:conservation-m}
\end{equation}
%%%%%%%%%%%%%%%%%%%%%%%%%%%%%%%%%%%%%%%%%%%%%%%%%%%%%%%%%%%%%%%%%%%%%%%%%%%%%%%%%%%%%%%%%%%%%%%%%
\begin{equation}
   \dot{\rho}_x + 3H(1+w)\rho_x = -Q \, .
\label{eq:conservation-x}
\end{equation}
%%%%%%%%%%%%%%%%%%%%%%%%%%%%%%%%%%%%%%%%%%%%%%%%%%%%%%%%%%%%%%%%%%%%%%%%%%%%%%%%%%%%%%%%%%%%%%%%%
Here $w = p_x/\rho_x < -1/3$ is the equation of state parameter of
DE, which for the sake of easiness we will assume to be constant.
Notice that the last two equations, together with the independent
conservation of $\rho_{b}$, guarantee the fulfilment of
(\ref{eq:conservation1}).

The next section  introduces, on a phenomenological basis,
appropriate expressions for $Q$ restricted by the conditions of
simplicity, smallness,  and positivity. The latter comes from the
result that a negative $Q$ would violate the second law of
thermodynamics \cite{secondlaw}; i.e., the energy transfer must go
from DE to DM, not the other way around, in conformity with the
findings of Refs. \cite{elcio, german}. Likewise, $Q$ must be
small (i.e., it should be bound by $Q/3H\rho_{m} < 1$); otherwise,
the DM could still dominate the expansion today.

A key quantity in interacting scenarios is the ratio between DM
and DE, $r = \rho_m/\rho_x$. In general, we expect it to
monotonically decrease with expansion. Equations
(\ref{eq:conservation-m}) and (\ref{eq:conservation-x}) imply
%%%%%%%%%%%%%%%%%%%%%%%%%%%%%%%%%%%%%%%%%%%%%%%%%%%%%%%%%%%%%%%%%%%%%%%%%
\begin{equation}
\frac{\dot{r}}{r} = 3Hw \, + \, (1+r) \, \frac{Q}{\rho_{x}} \,.
\label{eq:rdot}
\end{equation}
%%%%%%%%%%%%%%%%%%%%%%%%%%%%%%%%%%%%%%%%%%%%%%%%%%%%%%%%%%%%%%%%%%%%%%%%
Thus, the coincidence problem will be alleviated if $r$ decreases
with expansion slower than in the conventional $\Lambda$CDM model,
and much more so if for an extended expanse of time around the
present epoch the right-hand side of the last expression stays
close to zero. Obviously, for any of them  to occur, $Q$ must be
positive definite. In general,  $\dot r$ should  be negative at
all times. This leads to a further upper bound on $Q$:
%%%%%%%%%%%%%%%%%%%%%%%%%%%%%%%%%%%%%%%%%%%%%%%%%%%%%%%%%%%%%%%%%%%%%%%%%
\begin{equation}
Q \leq -3 H w \, \frac{\rho_{x}}{1+r} \, ,
 \label{eq:upperbound}
\end{equation}
%%%%%%%%%%%%%%%%%%%%%%%%%%%%%%%%%%%%%%%%%%%%%%%%%%%%%%%%%%%%%%%%%%%%%%%%
that may well be satisfied. Indeed, at early times one expects $r$
and $H$ to be much larger than their respective current values;
and at times later than the present one, one expects both of them
to be smaller. In general, if it were not so (i.e., if $\dot{r}$
were positive at late times), then the deceleration parameter, $q
= -1 -(\dot{H}/H^{2})$, some time in the future would evolve from
negative to positive values, conflicting with the second law of
thermodynamics \cite{grg-nd}. Thus, the bound  above is not at all
unreasonable.

The aim of this paper is twofold: first, to explore to what extent
some specific interactions terms (i.e., Q) alleviate the
coincidence problem and, second,  to investigate how accurate
measurements of $H(z)$ must  be to tell us, for the chosen {\em
As\"{a}tze}, whether  there really is an interaction in the dark
sector.

We believe this research is well justified in view of the progress
made in the recent years on the observational side. Indeed, it is
not an exaggeration to say that we are living in an exciting time
for cosmology. Currently, there are a number of ongoing surveys
and many more are planned -see Sec. 14 of Ref. \cite{li-li}. The
assured result is that the frontier of our knowledge on cosmology
will be pushed rather far away. In the next two decades,
measurements of $H(z)$ will be improved from the current accuracy
of roughly $12\%$ to better than $1\%$. This will allow one to
test cosmological models with unprecedented confidence.

The outline of the paper is as follows: Section II presents two
classes of interactions (I and II) and recalls the solutions to
the conservation equations for the interaction {\em Ans\"{a}tze}
adopted that have analytical solution. Section III investigates
whether these models really address the coincidence problem.
Section IV makes forecasts about detecting an interaction based on
measurements of the Hubble parameter. Section V summarizes our
findings and makes some final comments. As is usual, a subscript
zero denotes the current value of the corresponding quantity.

%%%%%%%%%%%%%%%%%%%%%%%%%%%%%%%%%%%%%%%%%%%%%%%%%%%%%%%%%%%%%%%%%%%%%%%%%%%%%%%%%%%
%%%%%%%%%%%%%%%%%%%%%%%%%%%%%%%%%%%%%%%%%%%%%%%%%%%%%%%%%%%%%%%%%%%%%%%%%%%%%%%%%%%
%%%%%%%%%%%%%%%%%%%%%%%%%%%%%%%%%%%%%%%%%%%%%%%%%%%%%%%%%%%%%%%%%%%%%%%%%%%%%%%%%%%
%%%%%%%%%%%%%%%%%%%%%%%%%%%%%%%%%%%%%%%%%%%%%%%%%%%%%%%%%%%%%%%%%%%%%%%%%%%%%%%%%%%
\section{Interacting models}
>From Eqs. (\ref{eq:conservation-m}) and (\ref{eq:conservation-x}),
$Q$ must be a function of the energy densities multiplied by a
quantity with units of inverse of time. For the latter we may take
the Hubble expansion rate; thus, $Q = Q(H\rho_{m}, H\rho_{x})$. By
power-law expanding last expression and retaining just the first
term, we arrive to the first class of interacting models:
%%%%%%%%%%%%%%%%%%%%%%%%%%%%%%%%%%%%%%%%%%%%%%%%%%%%%%%%%%%%%%%%%%%
%%%%%%%%%%%%%%%%%%%%%%%%%%%%%%%%%%%%%%%%%%%%%%%%%%%%%%%%%%%%%%%%%%%
\begin{equation}
 \mbox{I}: \;\;   Q_{1} = 3H (\epsilon_m \rho_m + \epsilon_x \rho_x) \, ,
 \label{eq:model1}
\end{equation}
%%%%%%%%%%%%%%%%%%%%%%%%%%%%%%%%%%%%%%%%%%%%%%%%%%%%%%%%%%%%%%%%%%%
%%%%%%%%%%%%%%%%%%%%%%%%%%%%%%%%%%%%%%%%%%%%%%%%%%%%%%%%%%%%%%%%%%%
where the parameters $\epsilon_m$ and $\epsilon_x$ are small
positive-semidefinite constants. This class, with $\epsilon_m =
\epsilon_x > 0$, leads a constant ratio, $r$, at late times, thus
solving the coincidence problem
\cite{Chimento2003a,Chimento2003b}.

As a further class we consider
%%%%%%%%%%%%%%%%%%%%%%%%%%%%%%%%%%%%%%%%%%%%%%%%%%%%%%%%%%%%%%%%%%%%%%%%%%%%%%%%
\begin{equation}
 \mbox{II}: \;\;   Q_{2} = 3 (\Gamma_m \rho_m + \Gamma_x \rho_x) \, ,
 \label{eq:model2}
\end{equation}
%%%%%%%%%%%%%%%%%%%%%%%%%%%%%%%%%%%%%%%%%%%%%%%%%%%%%%%%%%%%%%%%%%%%%%%%%%%%%%%%
where  $\Gamma_m$ and $\Gamma_x$ are non-negative, small, constant
time rates (by ``small" we mean smaller than $H_{0}$).
Cosmological models featuring this type of interaction in the dark
sector are motivated by similar models in reheating, curvaton
decay, and decay of DM into radiation \cite{Caldera-Cabral2009}.

Throughout this paper we will adopt the following specific {\em
Ans\"{a}tze} for the DM-DE interaction, Eqs. (\ref{eq:model1}) and
(\ref{eq:model2}):
%%%%%%%%%%%%%%%%%%%%%%%%%%%%%%%%%%%%%%%%%%%%%%%%%%%%%%%%%%%%%%%%%%%%%%%%%%%%%%%%%%%%%%%%%%%
\begin{equation}
 Q_{1a} = 3\epsilon H \rho_m \, ,
 \label{eq:Q1a}
\end{equation}
%%%%%%%%%%%%%%%%%%%%%%%%%%%%%%%%%%%%%%%%%%%%%%%%%%%%%%%%%%%%%%%%%%%%%%%%%%%%%%%%%%%%%%%%%%%
\begin{equation}
 Q_{1b} = 3\epsilon H \rho_x \, ,
 \label{eq:Q1b}
\end{equation}
%%%%%%%%%%%%%%%%%%%%%%%%%%%%%%%%%%%%%%%%%%%%%%%%%%%%%%%%%%%%%%%%%%%%%%%%%%%%%%%%%%%%%%%%%%%
\begin{equation}
 Q_{1c} = 3\epsilon H (\rho_m+\rho_x) \,  ,
 \label{eq:Q1c}
\end{equation}
%%%%%%%%%%%%%%%%%%%%%%%%%%%%%%%%%%%%%%%%%%%%%%%%%%%%%%%%%%%%%%%%%%%%%%%%%%%%%%%%%%%%%%%%%%%
\begin{equation}
 Q_{2a} = 3\Gamma \rho_m \, ,
 \label{eq:Q2a}
\end{equation}
%%%%%%%%%%%%%%%%%%%%%%%%%%%%%%%%%%%%%%%%%%%%%%%%%%%%%%%%%%%%%%%%%%%%%%%%%%%%%%%%%%%%%%%%%%%
\begin{equation}
 Q_{2b} = 3\Gamma \rho_x \, ,
 \label{eq:Q2b}
\end{equation}
%%%%%%%%%%%%%%%%%%%%%%%%%%%%%%%%%%%%%%%%%%%%%%%%%%%%%%%%%%%%%%%%%%%%%%%%%%%%%%%%%%%%%%%%%%%
\begin{equation}
 Q_{2c} = 3\Gamma (\rho_m+\rho_x) \, .
 \label{eq:Q2c}
\end{equation}
%%%%%%%%%%%%%%%%%%%%%%%%%%%%%%%%%%%%%%%%%%%%%%%%%%%%%%%%%%%%%%%%%%%%%%%%%%%%%%%%%%%%%%%%%%%

Models of type I explicitly depend on $H(z)$, while type II models
do not. The conservation equations for models of type I have
analytical solutions for all three cases considered (see, for
example, \cite{delCampo2011}). We next recall them.

%%%%%%%%%%%%%%%%%%%%%%%%%%%%%%%%%%%%%%%%%%%%%%%%%%%%%%%%%%%%%%%%%%%%%%%%%%%%%%%%%%%%%%%%%%%
%%%%%%%%%%%%%%%%%%%%%%%%%%%%%%%%%%%%%%%%%%%%%%%%%%%%%%%%%%%%%%%%%%%%%%%%%%%%%%%%%%%%%%%%%%%
\subsection{{\em Ansatz} $Q_{1a}$}
%
%%%%%%%%%%%%%%%%%%%%%%%%%%%%%%%%%%%%%%%%%%%%%%%%%%%%%%%%%%%%%%%%%%%%%%%%%%%%%%%%%%%%%%%%%%%
\begin{equation}
 \rho_m = \rho_{m0} \; x^{3(1-\epsilon)} ,
\end{equation}
%%%%%%%%%%%%%%%%%%%%%%%%%%%%%%%%%%%%%%%%%%%%%%%%%%%%%%%%%%%%%%%%%%%%%%%%%%%%%%%%%%%%%%%%%%%
\begin{equation}
 \rho_x = \rho_{x0} \; x^{3(1+w)} + \left( \frac{\epsilon}{\epsilon+w} \right) \rho_{m0} [x^{3(1+w)} - x^{3(1-\epsilon)}] ,
\end{equation}
%%%%%%%%%%%%%%%%%%%%%%%%%%%%%%%%%%%%%%%%%%%%%%%%%%%%%%%%%%%%%%%%%%%%%%%%%%%%%%%%%%%%%%%%%%%
where $x = 1+z = \frac{a_0}{a}$ with $a$ the scale factor of the
Friedmann-Robertson-Walker metric.

\subsection{{\em Ansatz} $Q_{1b}$}
%
%%%%%%%%%%%%%%%%%%%%%%%%%%%%%%%%%%%%%%%%%%%%%%%%%%%%%%%%%%%%%%%%%%%%%%%%%%%%%%%%%%%%%%%%%%%
\begin{equation}
 \rho_m = \rho_{m0} \; x^{3} + \left( \frac{\epsilon}{\epsilon+w} \right)
 \rho_{x0} [1 - x^{3(w+\epsilon)}]x^3 \, ,
\end{equation}
%%%%%%%%%%%%%%%%%%%%%%%%%%%%%%%%%%%%%%%%%%%%%%%%%%%%%%%%%%%%%%%%%%%%%%%%%%%%%%%%%%%%%%%%%%%
\begin{equation}
 \rho_x = \rho_{x0} \; x^{3(1+w+\epsilon)} \, .
\end{equation}
%%%%%%%%%%%%%%%%%%%%%%%%%%%%%%%%%%%%%%%%%%%%%%%%%%%%%%%%%%%%%%%%%%%%%%%%%%%%%%%%%%%%%%%%%%%

\subsection{{\em Ansatz} $Q_{1c}$}
%
%%%%%%%%%%%%%%%%%%%%%%%%%%%%%%%%%%%%%%%%%%%%%%%%%%%%%%%%%%%%%%%%%%%%%%%%%%%%%%%%%%%%%%%%%%%
\begin{equation}
 \rho_m = C_1x^{\gamma_1} + C_2x^{\gamma_2} \, ,
\end{equation}
%%%%%%%%%%%%%%%%%%%%%%%%%%%%%%%%%%%%%%%%%%%%%%%%%%%%%%%%%%%%%%%%%%%%%%%%%%%%%%%%%%%%%%%%%%%
\begin{equation}
 \rho_x = \frac{1}{2\epsilon} \left[ -C_1(A+B) + C_2(B-A)x^{-3B} \right] x^{\gamma_1}
 \, ,
 \label{eq:rho_x-Q1c}
\end{equation}
%%%%%%%%%%%%%%%%%%%%%%%%%%%%%%%%%%%%%%%%%%%%%%%%%%%%%%%%%%%%%%%%%%%%%%%%%%%%%%%%%%%%%%%%%%%
where $C_{1,2}$, $\gamma_{1,2}$, $A$ and $B$ are defined below:
%%%%%%%%%%%%%%%%%%%%%%%%%%%%%%%%%%%%%%%%%%%%%%%%%%%%%%%%%%%%%%%%%%%%%%%%%%%%%%%%%%%%%%%%%%%
\begin{equation}
 \gamma_{1} = \frac{3}{2}(2+w+B) \, , \; \quad
 \gamma_{2} = \frac{3}{2}(2+w-B) \, ,
\end{equation}
%%%%%%%%%%%%%%%%%%%%%%%%%%%%%%%%%%%%%%%%%%%%%%%%%%%%%%%%%%%%%%%%%%%%%%%%%%%%%%%%%%%%%%%%%%%
\begin{equation}
 C_1 = \frac{1}{2B} [(B-A)\rho_{m0} - 2\epsilon\rho_{x0}] , \;
 \quad
 C_2 = \frac{1}{2B} [(B+A)\rho_{m0} + 2\epsilon\rho_{x0}] ,
\end{equation}
%%%%%%%%%%%%%%%%%%%%%%%%%%%%%%%%%%%%%%%%%%%%%%%%%%%%%%%%%%%%%%%%%%%%%%%%%%%%%%%%%%%%%%%%%%%
\begin{equation}
 A = w + 2\epsilon , \; \quad
 B = \sqrt{w(w+4\epsilon)}\,  .
\end{equation}
%%%%%%%%%%%%%%%%%%%%%%%%%%%%%%%%%%%%%%%%%%%%%%%%%%%%%%%%%%%%%%%%%%%%%%%%%%%%%%%%%%%%%%%%%%%

Models of type II$a$ and II$b$ have an analytical solution for
just one of the conservation equations (for $\rho_m$ in case II$a$
and for $\rho_x$ in case II$b$, see \cite{Caldera-Cabral2009}),
while model II$c$ [Eq. (\ref{eq:Q2c})] has no analytical solution.
In any case, they are to be solved numerically.

Table \ref{table_epsilon} recalls some of the observational
constraints of type I models -- see Refs. \cite{He2011,
Basilakos2009, Costa2010}. To the best of our knowledge no
observational constraints for model II are known at this stage.

%%%%%%%%%%%%%%%%%%%%%%%%%%%%%%%%%%%%%%%%%%%%%%%%%%%%%%%%%%%%%%%%%%%%%%%%%%%%%%%%%%%%%%%%%%%%%%%%%%%%%%%%%%%%%%%%%%%%%%%%%%%%%%%%%%%%%%%
%%%%%%%%%%%%%%%%%%%%%%%%%%%%%%%%%%%%%%%%%%%%%%%%%%%%%%%%%%%%%%%%%%%%%%%%%%%%%%%%%%%%%%%%%%%%%%%%%%%%%%%%%%%%%%%%%%%%%%%%%%%%%%%%%%%%%%%
\begin{table*}[t]
  \centering
  \begin{tabular}{*{4}{c}} \toprule
     Reference &   $Q_{1a}$ &   $Q_{1b}$ &   $Q_{1c}$ \\
    \hline
     \cite{He2011} $(w<-1)$ $\;\;$ & $\;\;$ $\epsilon = 0.0006^{+0.0006}_{-0.0005}$ $\;\;$ & $\epsilon = 0.024^{+0.034}_{-0.027}$ & $\epsilon = 0.0006^{+0.0005}_{-0.0006}$  \\
     \cite{He2011} $(w>-1)$ $\;\;$ & $\;\;$ None $\;\;$ & $\;\;$ $\epsilon = -0.003^{+0.017}_{-0.024}$ $\;\;$ & $\;\;$None  \\
     \cite{Basilakos2009} $ (w=-1)$ $\;\;$ & $\;\;$ $ \epsilon = 0.002 \pm 0.001 $ $\;\;$ & $\;\;$ None $\;\;$ & $\;$ None \\
     \cite{Costa2010} $(w=-1)$ $\;\;$ & $\;\;$ $\epsilon = 0.01 \pm 0.01 $ $\;\;$ & $\;\;$ None $\;\;$ & $\;$ None \\
    \hline
    \hline
  \end{tabular}
  \caption{Observational constraints on $\epsilon$ for type I models.}
  \label{table_epsilon}
\end{table*}
%%%%%%%%%%%%%%%%%%%%%%%%%%%%%%%%%%%%%%%%%%%%%%%%%%%%%%%%%%%%%%%%%%%%%%%%%%%%%%%%%%%%%%%%%%%%%%%%%%%%%%%%%%%%%%%%%%%%%%%%%%%%%%%%%%%%%%%%%
%%%%%%%%%%%%%%%%%%%%%%%%%%%%%%%%%%%%%%%%%%%%%%%%%%%%%%%%%%%%%%%%%%%%%%%%%%%%%%%%%%%%%%%%%%%%%%%%%%%%%%%%%%%%%%%%%%%%%%%%%%%%%%%%%%%%%%%

%%%%%%%%%%%%%%%%%%%%%%%%%%%%%%%%%%%%%%%%%%%%%%%%%%%%%%%%%%%%%%%%%%%%%%%%%%%%%%%%%%%%%%%%%%%%%%%%%%%%%%%%%
%%%%%%%%%%%%%%%%%%%%%%%%%%%%%%%%%%%%%%%%%%%%%%%%%%%%%%%%%%%%%%%%%%%%%%%%%%%%%%%%%%%%%%%%%%%%%%%%%%%%%%%%%
%%%%%%%%%%%%%%%%%%%%%%%%%%%%%%%%%%%%%%%%%%%%%%%%%%%%%%%%%%%%%%%%%%%%%%%%%%%%%%%%%%%%%%%%%%%%%%%%%%%%%%%%%
\section{Alleviating the coincidence problem}
As said above, an interaction in the dark sector, aside from being
an open theoretical option, provides an interesting possibility to
address the coincidence problem. This section focus on how the
interacting {\em As\"{a}tze}  given by Eqs.
(\ref{eq:Q1a})-(\ref{eq:Q2c}) fare with that issue by determining
(analytically for class I models and numerically for class II
models) the evolution of the ratio $r = \frac{\rho_m}{\rho_x}$,
whose present day value $r_0$ is of the order of the unity. The
coincidence problem will be alleviated if $r \simeq r_0$ for a
substantially longer period about the present time than in the
$\Lambda$CDM model. This has been investigated in some extent in
references \cite{delCampo2008}-\cite{He2009}.

\begin{figure*}[htbp]
    \centering
        \includegraphics[width=0.31\linewidth]{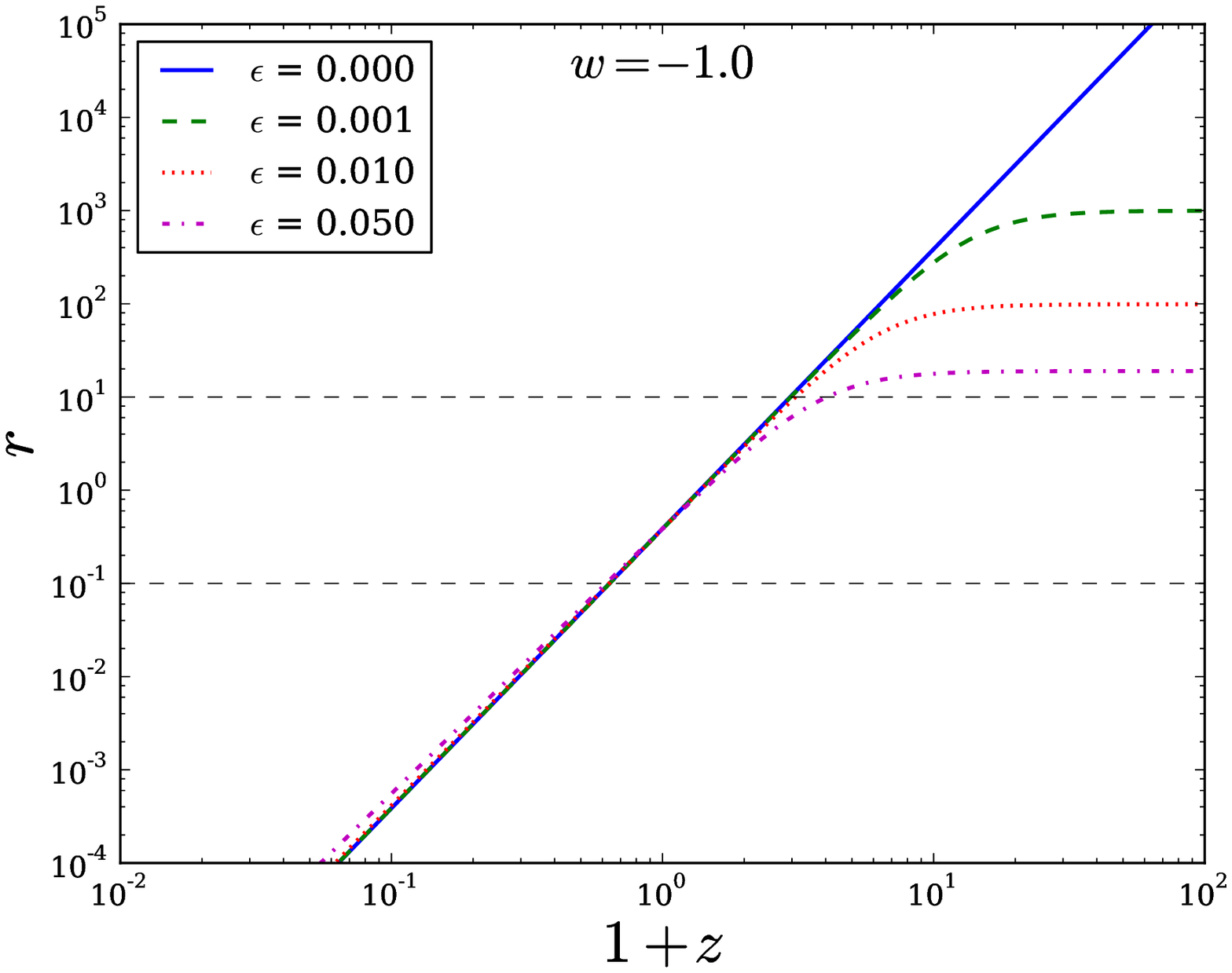}
        \includegraphics[width=0.31\linewidth]{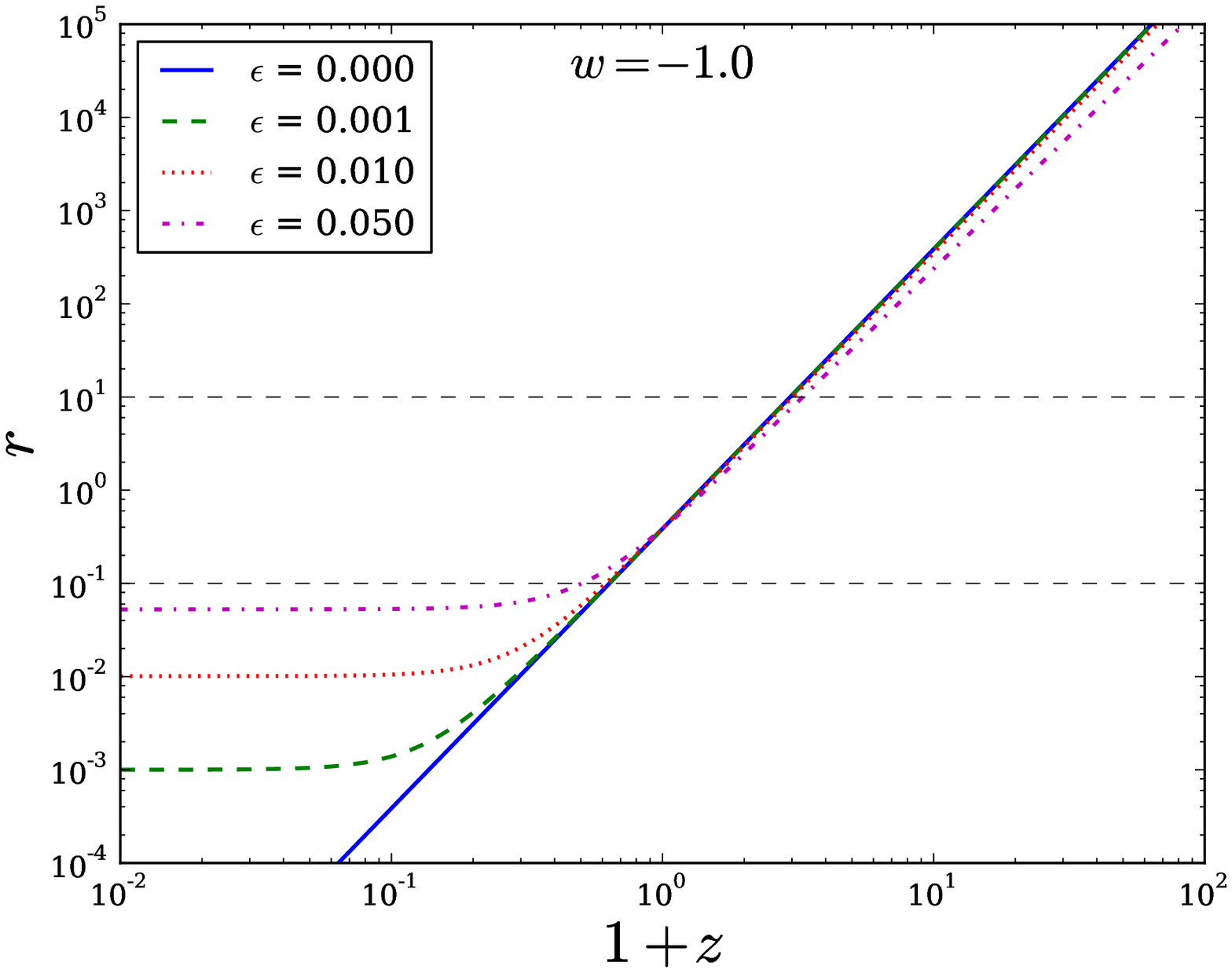}
        \includegraphics[width=0.31\linewidth]{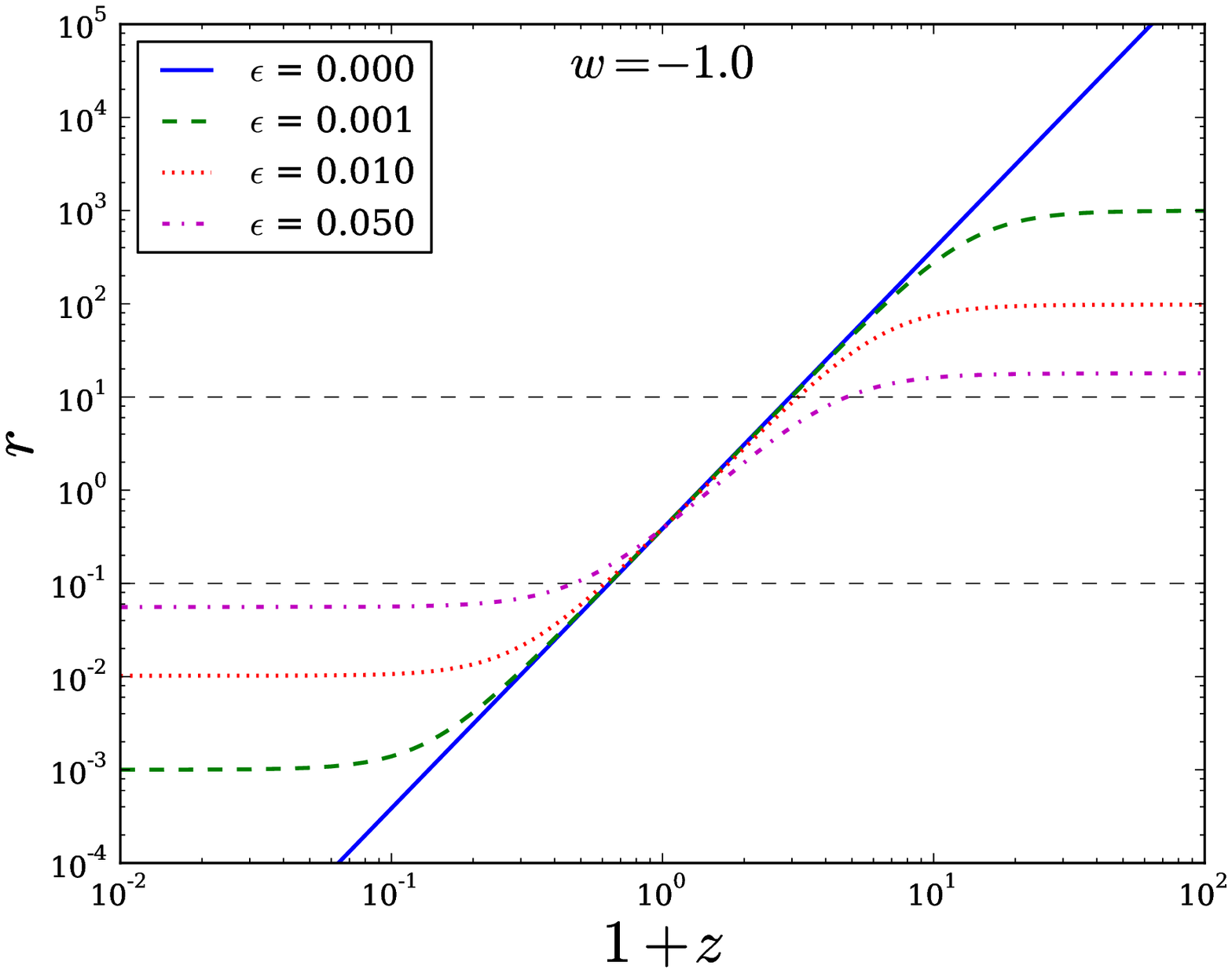}
    \caption{Plots of $r$ as a function of $1+z$ for {\em As\"{a}tze} $Q_{1a}$ (left panel), $Q_{1b}$
    (middle panel) and $Q_{1c}$ (right panel). The dashed horizontal lines correspond to
    $r=0.1$ and $r=10$. As can be seen from Eq. (\ref{eq:rho_x-Q1c}), model $Q_{1c}$ has
    an apparent divergence for $\epsilon = 0$ (analytic computation of the limit
    $\epsilon \rightarrow 0$ shows that it really exists), so we put a very small
    value for $\epsilon$ that represents the $\Lambda$CDM for comparison purposes. In
    plotting the graphs we  used $H_0 = 67.4$, $\Omega_b = 0.049$,
    $\Omega_m = 0.265$ and $\Omega_x = 1 - \Omega_m - \Omega_b$.}
    \label{fig:CP_Q1_w-10}
%\end{figure*}
%\begin{figure*}[htbp]
    \centering
        \includegraphics[width=0.31\linewidth]{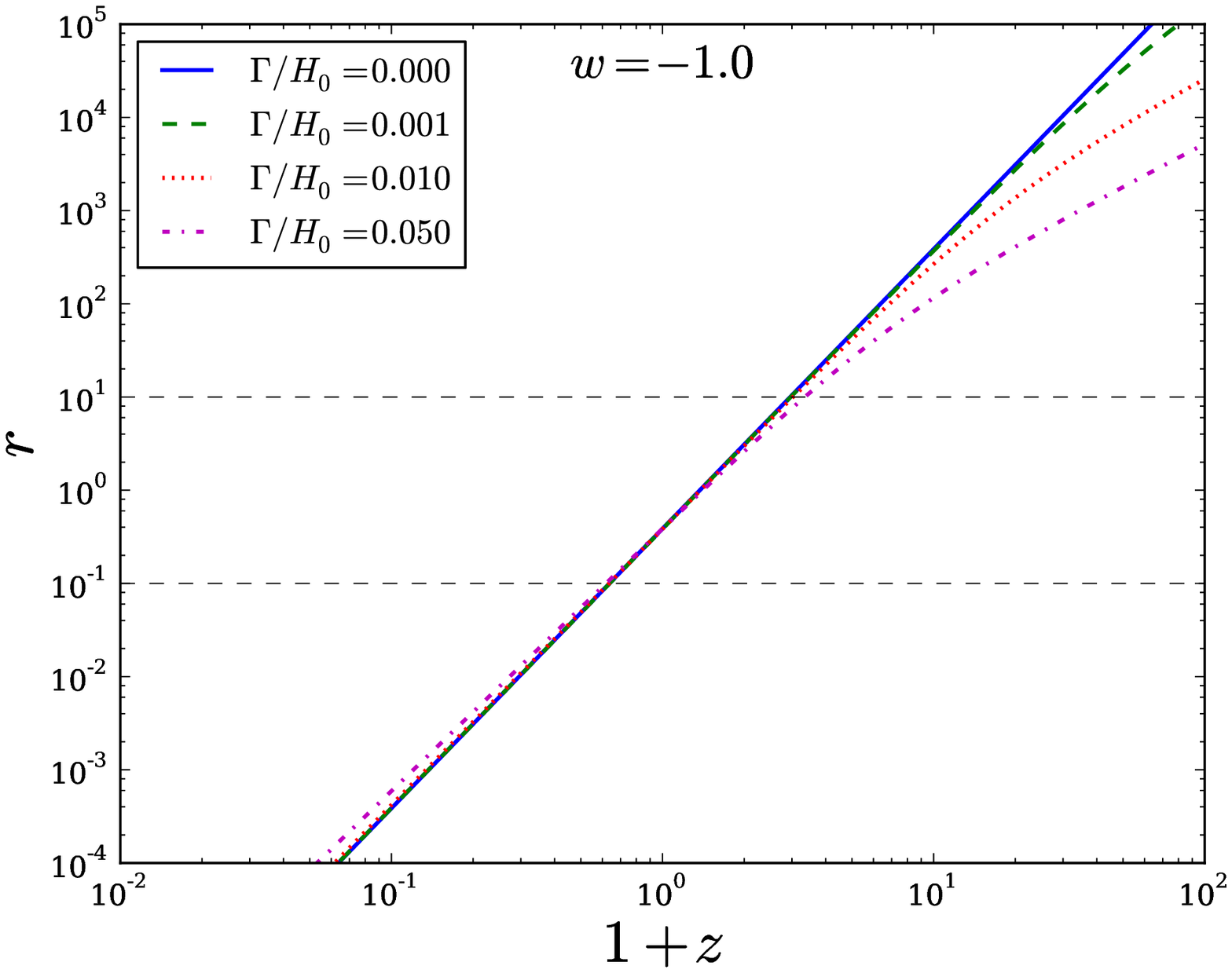}
        \includegraphics[width=0.31\linewidth]{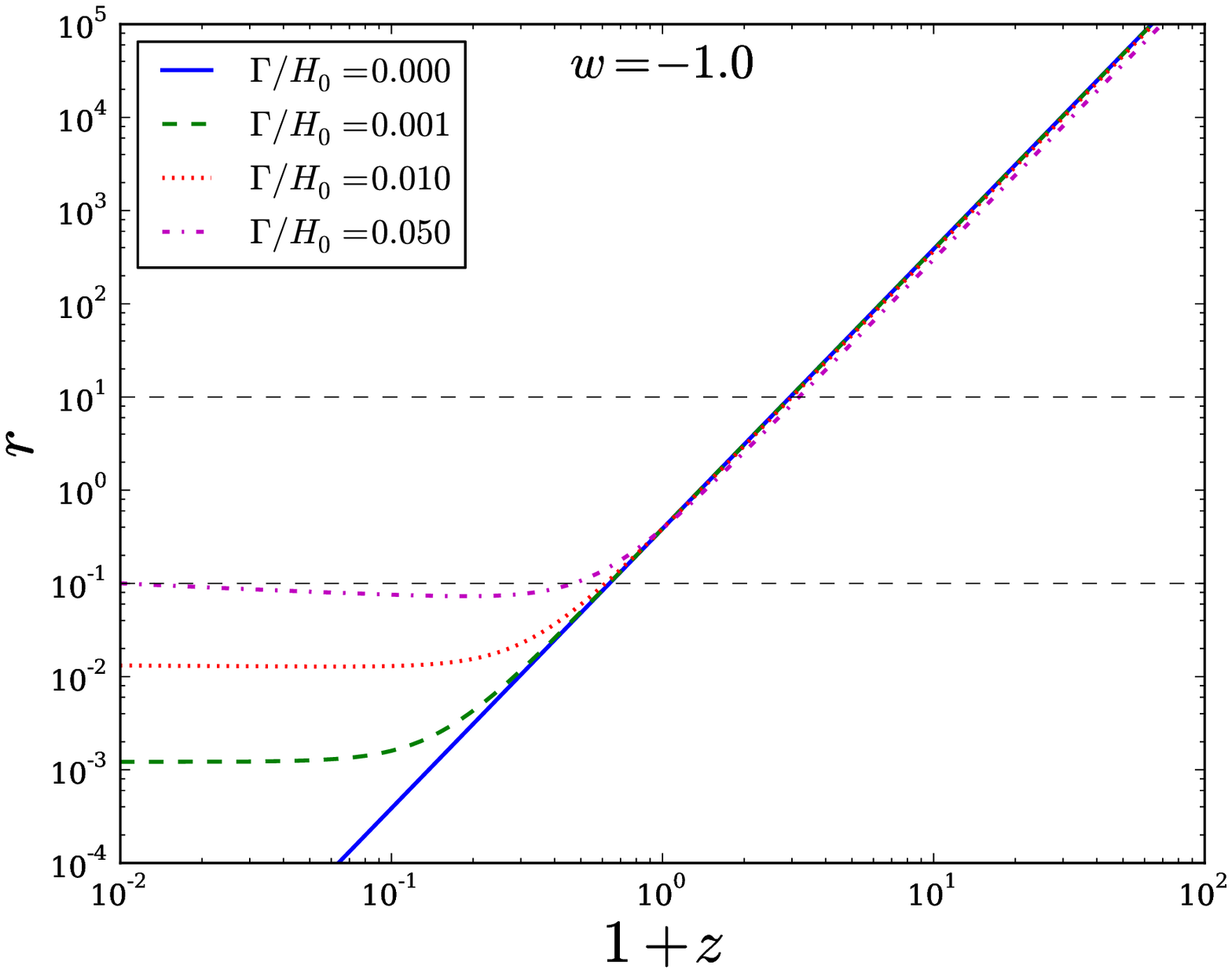}
        \includegraphics[width=0.31\linewidth]{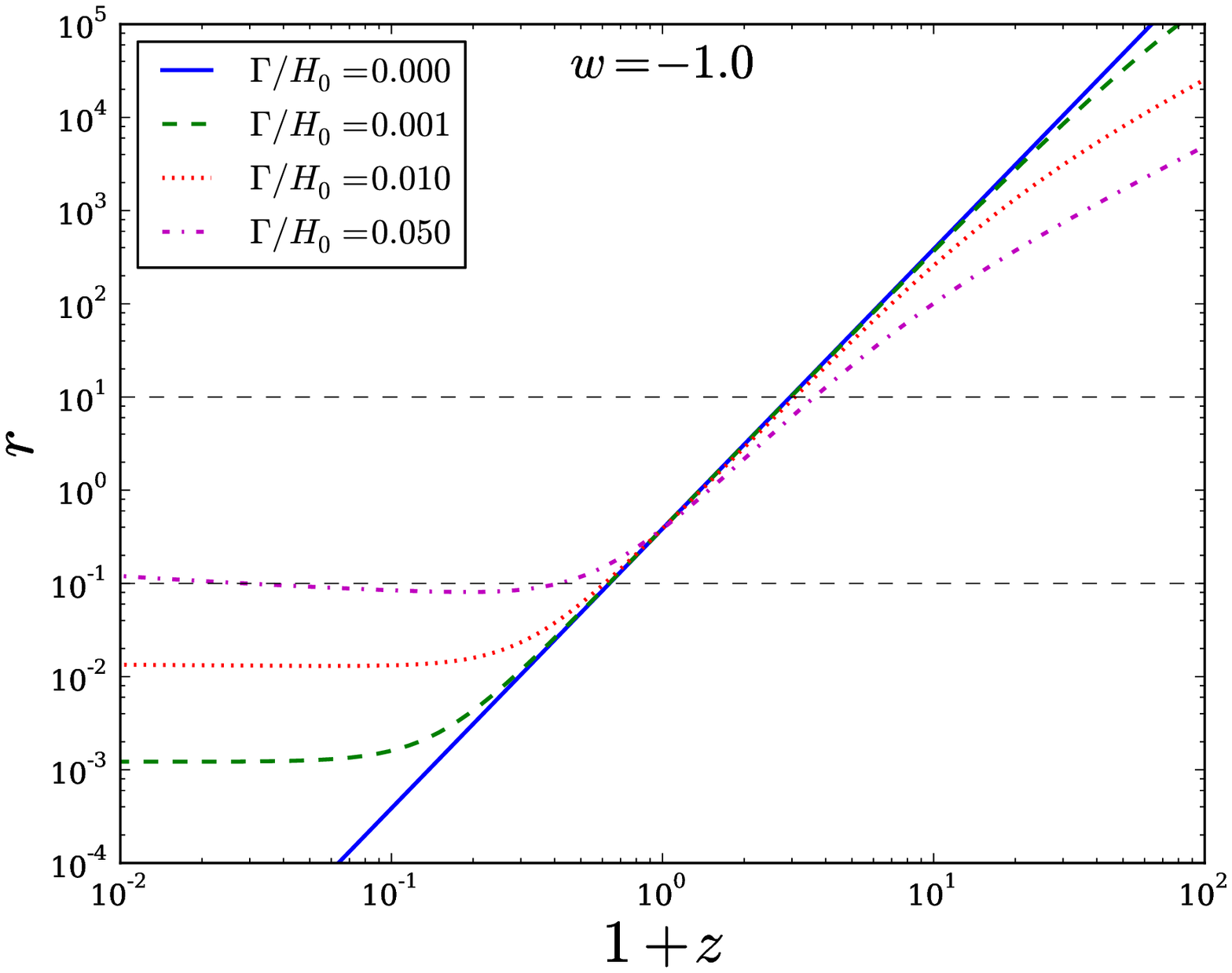}
    \caption{The same as Fig. \ref{fig:CP_Q1_w-10} but for the {\em Ans\"{a}tze} $Q_{2a}$ (left panel),
    $Q_{2b}$ (middle panel), and $Q_{2c}$ (right panel).}
    \label{fig:CP_Q2_w-10}
\end{figure*}

Figures \ref{fig:CP_Q1_w-10} and \ref{fig:CP_Q2_w-10} plot $r \,$
vs $\, 1+z$ (with $w=-1.0$) for models of type I and II,
respectively. They  result from  integrating the conservation
equations (\ref{eq:conservation-m}) and (\ref{eq:conservation-x}),
in each case using the corresponding expression for $Q$. The
values of $\epsilon$  were selected to span current constraints
(see Table \ref{table_epsilon}) and those corresponding to
$\Gamma/H_0$ were set to the same set of values.

Throughout this paper we take $H_0 = 67.4$, $\Omega_{b0}= 0.049$,
$\Omega_{m0} = 0.265$, and $\Omega_{x0} = 1 - \Omega_{m0} -
\Omega_{b0}$, the central values reported by the Planck
Collaboration \cite{PlanckCosmo2013}, and, accordingly, we use
$r_0 = 0.386$. As usual, the $\Omega$ quantities denote
fractional densities.

%%%%%%%%%%%%%%%%%%%%%%%%%%%%%%%%%%%%%%%%%%%%%%%%%%%%%%%%%%%%%%%%%%%%%%%%%%%%%%%%%%%%%%%%%%%%%%%%%%%%%%%%%
%%%%%%%%%%%%%%%%%%%%%%%%%%%%%%%%%%%%%%%%%%%%%%%%%%%%%%%%%%%%%%%%%%%%%%%%%%%%%%%%%%%%%%%%%%%%%%%%%%%%%%%%%
%%%%%%%%%%%%%%%%%%%%%%%%%%%%%%%%%%%%%%%%%%%%%%%%%%%%%%%%%%%%%%%%%%%%%%%%%%%%%%%%%%%%%%%%%%%%%%%%%%%%%%%%%
%%%%%%%%%%%%%%%%%%%%%%%%%%%%%%%%%%%%%%%%%%%%%%%%%%%%%%%%%%%%%%%%%%%%%%%%%%%%%%%%%%%%%%%%%%%%%%%%%%%%%%%%%

>From Fig. \ref{fig:CP_Q1_w-10} [that corresponds to models of type
I, directly depending on $H(z)$], it is seen that $r$ deviates
from the $\Lambda$CDM model at different redshift intervals
depending on the form of the interaction. When $Q$ depends on
$\rho_m$, $r$ deviates in the  matter-dominated region (left
panel). When $Q \propto \rho_x$ the deviation occurs mostly in the
future ($z<0$), where dark energy prevails (middle panel).
Finally, when $Q \propto (\rho_{m} +\rho_{x})$, right panel, the
deviations occur both in the past and in the future. Although the
curves for $r$ in the interacting model look almost the same as
for the $\Lambda$CDM model in the vicinity of the present age
(horizontal dashed lines in Fig. \ref{fig:CP_Q1_w-10}), they limit
the range which $r$ varies in the past (left panel), future
(middle panel), or both past and future (right panel). While this
does not help to clarify why $r$ is of the order of unity today,
it narrows the range $r$ spans. For instance, model $Q_{1c}$
(right panel of Fig. \ref{fig:CP_Q1_w-10}) limits $r$ in the range
$10^{-3}$ - $10^3$ (for $\epsilon = 0.001$), thus alleviating the
coincidence problem. Similarly, models $Q_{1a}$ (left panel) and
$Q_{1b}$ (middle panel) narrow the allowed range of $r$ in the
past and future, respectively. In this sense, all three {\em
As\"{a}tze} alleviate  the coincidence problem for positive values
of $\epsilon$ (negative values only worsen the problem). More
specifically, from the middle panel of  said figure we learn that
at the very far future, $z = -0.99$, $r$ takes values not lower
than $10^{-3}$ for the interacting models displayed in that panel,
but for the $\Lambda$CDM it goes down to $10^{-6}$. Like comments
apply to the right-hand panel.

As is well known, the contribution of dark energy (with constant
$w$) at high redshifts is substantially constrained \cite{bean,
xia-viel}. Combining a variety of observational data, Xia and Viel
\cite{xia-viel} found that the fractional density of dark energy
at the time of last scattering surface should not exceed $1.4
\times 10^{-3}$ (at $95\%$ confidence level). This implies that in
the case of {\em Ans\"{a}tze} $Q_{1a}$ and $Q_{1c}$ (left and
right panels of Fig. \ref{fig:CP_Q1_w-10}, respectively)
$\epsilon$ should be of the order of $10^{-3}$ or less.

Curiously enough, {\em Ans\"{a}tze} $Q_{1b}$ and $Q_{2b}$ lead to
very similar evolutions of $r$, particularly for $z<0$. Figure
\ref{fig:CP_Q2_w-10} [corresponding to models of type II, which do
not  directly depend on $H(z)$] shows that the {\em Ansatz}
$Q_{2a}$ barely alleviates the coincidence problem (left panel),
for it does not limit $r$  in either the past or future. The other
two {\em Ans\"{a}tze} (middle and right panels) do alleviate it,
limiting $r$ in the future (in a similar way of the {\em
Ans\"{a}tze} of Fig. \ref{fig:CP_Q1_w-10}). We also notice the
same features mentioned above for models of type I (Fig.
\ref{fig:CP_Q1_w-10}), explaining the deviations of $r$ from
$\Lambda$CDM according to the dependencies of $Q$ ($\rho_m$,
$\rho_x$ or $\rho_m + \rho_x$). By contrast to
\cite{Caldera-Cabral2009}, we see that interactions of type II
({\em Ansatz} $Q_{2c}$ with $w = -1$, right panel of Fig.
\ref{fig:CP_Q2_w-10}) are well behaved at all times and also
address the coincidence problem, as long as $\Gamma/H_0 \lesssim
0.01$, when $r$ is also well behaved up to at least  $1+z =
10^{-15}$. The same holds true for the claim in
\cite{Caldera-Cabral2009} that in the case of the {\em Ansatz}
$Q_{2b}$ the Universe would be dominated by DM in the future: If
$\Gamma/H_0 \lesssim 0.01$, we find that DE dominates until at
least $1+z = 10^{-15}$.

>From Figs. \ref{fig:CP_Q1_w-10} and \ref{fig:CP_Q2_w-10} we
conclude that for $\epsilon \simeq \Gamma /H_0$ the evolution of
$r$ does not greatly differ, particularly between the cases
$Q_{1b}$ and $Q_{2b}$. This may indicate that the constraints on
$\Gamma /H_0$ will be of the same order of magnitude as those of
$\epsilon$ for corresponding models.

Altogether, we can say that the {\em Ans\"{a}tze} that most
alleviate the coincidence problem, and do not conflict with
present constraints on the abundance of dark energy at early
times, are $Q_{1b}$, $Q_{2b}$, and $Q_{2c}$.

\begin{figure*}[htbp]
    \centering
        \includegraphics[width=0.31\linewidth]{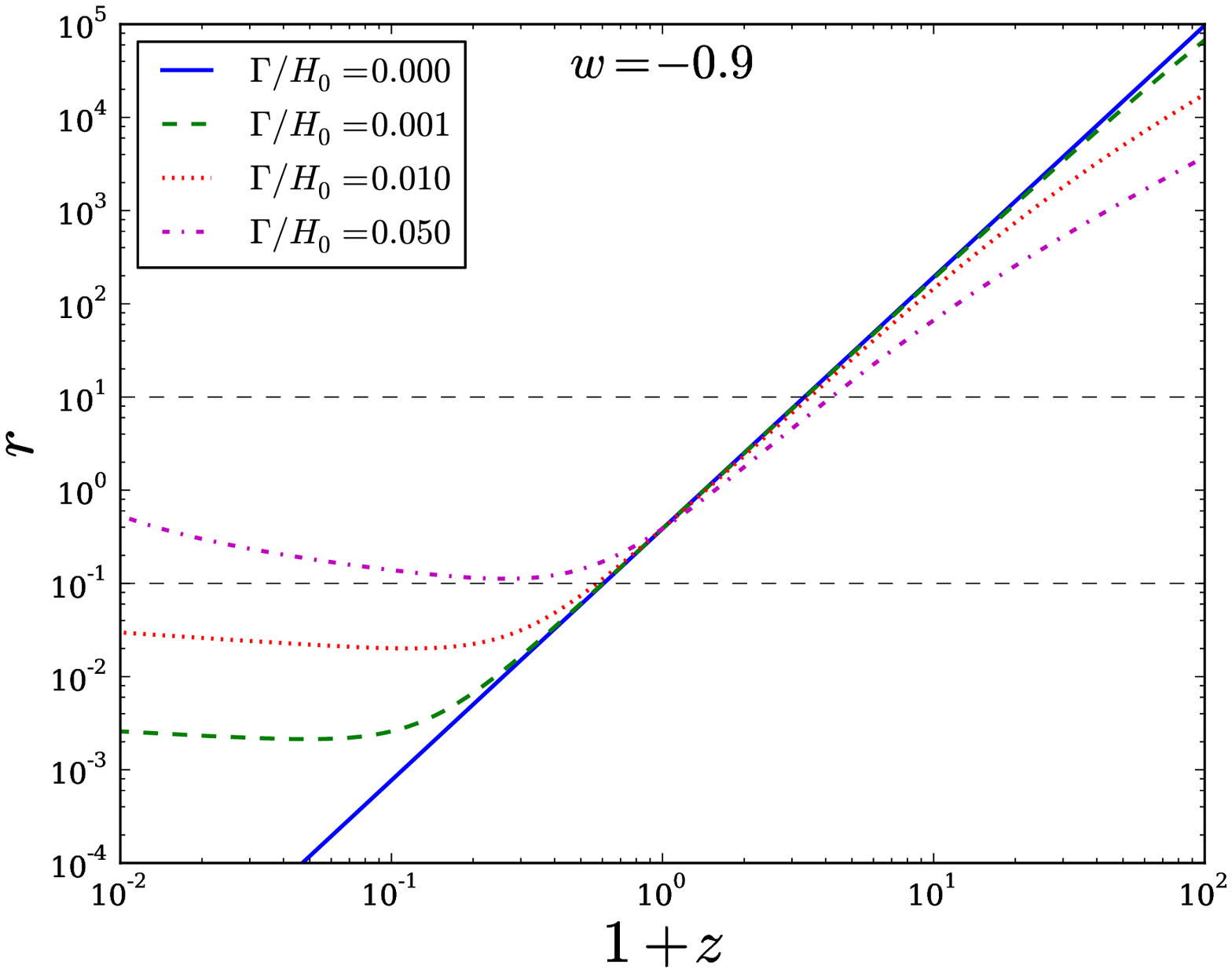}
        \includegraphics[width=0.31\linewidth]{plot_matter_to_DE_ratio_log_w-10_Q2c.eps}
        \includegraphics[width=0.31\linewidth]{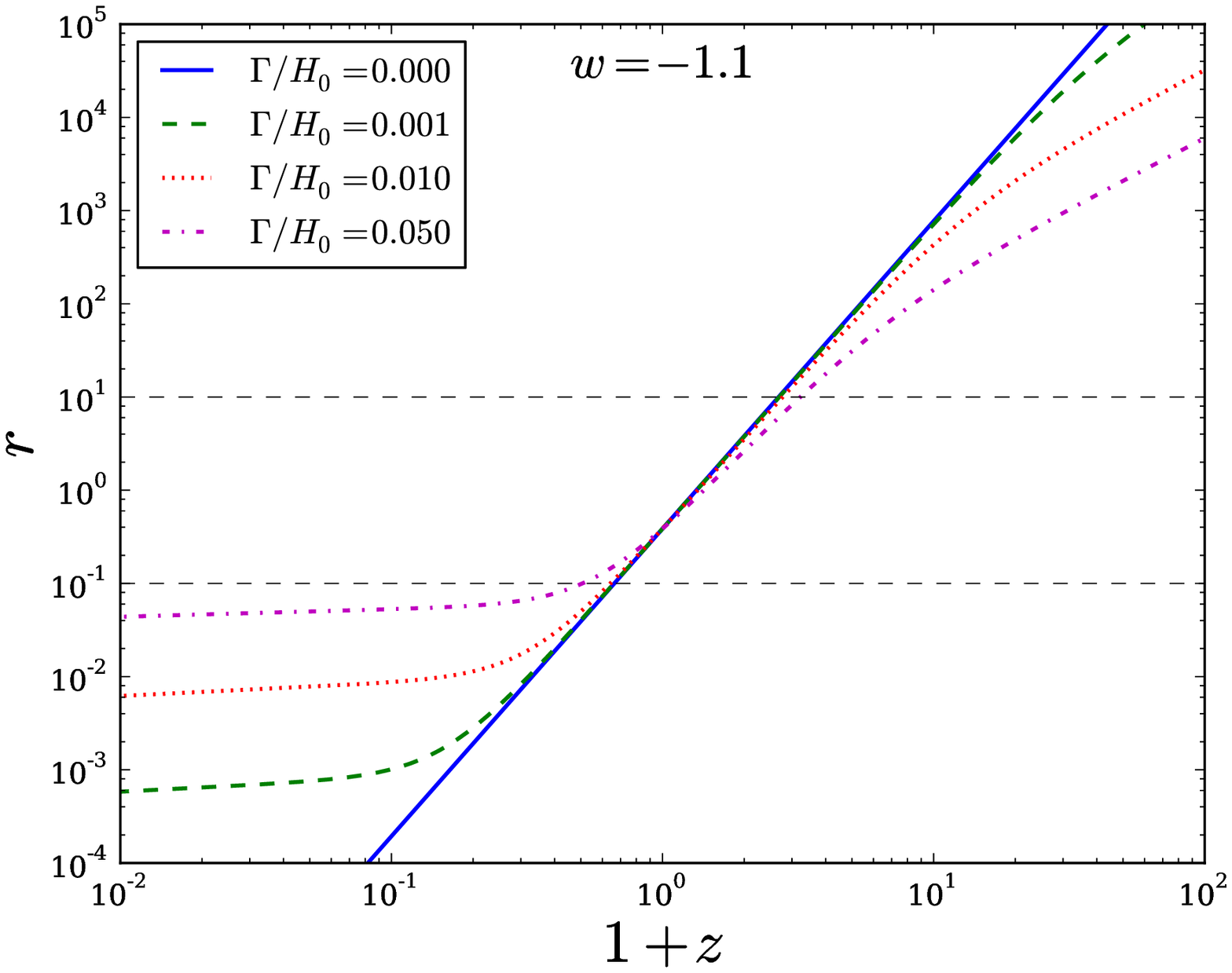}
    \caption{Dependence on $w$ for {\em Ansatz} $Q_{2c}$ when $w=-0.9$ (left panel), $w=-1.0$
    (middle panel) and $w=-1.1$ (right panel). In
    plotting the graphs we  used $H_0 = 67.4$, $\Omega_b = 0.049$,
    $\Omega_m = 0.265$, and $\Omega_x = 1 - \Omega_m - \Omega_b$.}
    \label{fig:CP_Q2c_ws}
\end{figure*}

As an example of dependency of the densities ratio $r$ with the
equation of state parameter $w$, Fig. \ref{fig:CP_Q2c_ws} shows
plots for three values of $w$ for the {\em Ansatz}  $Q_{2c}$. It
is seen that  $w <-1$ (phantom values) tends to worsen the
coincidence problem, while for $w > -1$ (quintessence values) it
gets alleviated. It should be recalled that, in addition, phantom
models face severe problems regarding quantum stability
\cite{hoffman,cline} and violate the second law of thermodynamics
\cite{grg-nd}. Likewise, we wish to emphasize that in order not to
run into negative $\rho_{x}$ values, $\Gamma/H_{0}$ must be lower
than $ \sim \, 10^{-4}$ and $10^{-2}$ for $w = -0.9$ and $w = -1$,
respectively.

%%%%%%%%%%%%%%%%%%%%%%%%%%%%%%%%%%%%%%%%%%%%%%%%%%%%%%%%%%%%%%%%%%%%%%%%%%%%%%%%%%%%%%%%%%%%%%%%%%%%%%%%%%%%
%%%%%%%%%%%%%%%%%%%%%%%%%%%%%%%%%%%%%%%%%%%%%%%%%%%%%%%%%%%%%%%%%%%%%%%%%%%%%%%%%%%%%%%%%%%%%%%%%%%%%%%%%%%%
%%%%%%%%%%%%%%%%%%%%%%%%%%%%%%%%%%%%%%%%%%%%%%%%%%%%%%%%%%%%%%%%%%%%%%%%%%%%%%%%%%%%%%%%%%%%%%%%%%%%%%%%%%%%
%%%%%%%%%%%%%%%%%%%%%%%%%%%%%%%%%%%%%%%%%%%%%%%%%%%%%%%%%%%%%%%%%%%%%%%%%%%%%%%%%%%%%%%%%%%%%%%%%%%%%%%%%%%%

%%%%%%%%%%%%%%%%%%%%%%%%%%%%%%%%%%%%%%%%%%%%%%%%%%%%%%%%%%%%%%%%%%%%%%%%%%%%%%%%%%%%%%%%%%%%%%%%%%%%%%%%%%%%
%%%%%%%%%%%%%%%%%%%%%%%%%%%%%%%%%%%%%%%%%%%%%%%%%%%%%%%%%%%%%%%%%%%%%%%%%%%%%%%%%%%%%%%%%%%%%%%%%%%%%%%%%%%%
%%%%%%%%%%%%%%%%%%%%%%%%%%%%%%%%%%%%%%%%%%%%%%%%%%%%%%%%%%%%%%%%%%%%%%%%%%%%%%%%%%%%%%%%%%%%%%%%%%%%%%%%%%%%
\section{Forecasts for detecting an interaction through $H(z)$ data}
In this section, we investigate the possibility of detecting an
interaction, i.e., whether DE and DM interact also
nongravitationally with each other, by using $H(z)$ data only. In
\cite{Ferreira2013} it was investigated for the specific {\em
Ansatz} $Q_{1a}$ with $w=-1$. Here we extend that analysis for the
models described in Sect. II, for constant $w$.

The error associated in measuring $H(z, \epsilon)$ propagates
according to
%%%%%%%%%%%%%%%%%%%%%%%%%%%%%%%%%%%%%%%%%%%%%%%%%%%%%%%%%%%%%%%%%%%%%%%%%%%%%%%%%%%%%%%%%%%%%%%%%%%%%
\begin{equation}
  \delta H^2 = \left( \frac{\partial H}{\partial \epsilon } \right)^2 \delta \epsilon^2
  \, ,
  \label{eq:err}
\end{equation}
%%%%%%%%%%%%%%%%%%%%%%%%%%%%%%%%%%%%%%%%%%%%%%%%%%%%%%%%%%%%%%%%%%%%%%%%%%%%%%%%%%%%%%%%%%%%%%%%%%%%%
and to an analogous expression for $H(z, \Gamma)$. Thus, with the
help of the last equation, one can calculate how the error in
$H(z)$ data propagates to $\epsilon$ (and correspondingly to
$\Gamma/H_{0}$) as a function of redshift. To allow a direct and
unambiguous forecast we compute the average of the relative H(z)
error, $\delta H/H$, over redshift [Eq. 5 of \cite{Ferreira2013}]:
%%%%%%%%%%%%%%%%%%%%%%%%%%%%%%%%%%%%%%%%%%%%%%%%%%%%%%%%%%%%%%%%%%%%%%%%%%%%%%%%%%%%%%%%%%%%%%%%%%%%%%
\begin{equation}
 \left< \frac{\delta H}{H} \right> = \left[ \frac{1}{z_f - z_i} \int_{z_i}^{z_f} \frac{1}{H}
 \left| \frac{\partial H}{\partial \epsilon} \right| dz \right] \delta \epsilon \;,
  \label{eqdeltaH}
\end{equation}
%%%%%%%%%%%%%%%%%%%%%%%%%%%%%%%%%%%%%%%%%%%%%%%%%%%%%%%%%%%%%%%%%%%%%%%%%%%%%%%%%%%%%%%%%%%%%%%%%%%%%%%
where [$z_i$, $z_f$] is the redshift interval in which the data
happen to fall. A corresponding expression holds for the case of
models belonging  to class II. Here we take $z_i = 0.07$ and $z_f
= 2.3$, that fix the interval  of the most  recent data - see
Table 1 in Ref. \cite{Farooq2013}. By calculating the average over
redshift, we allow a direct estimate for the required accuracy in
$H(z)$ observations to detect a given interaction. The quantities
$\partial H/ \partial \epsilon$  were computed analytically; the
quantities $\partial H/
\partial \Gamma$, numerically. Equation (\ref{eqdeltaH}) (and
their corresponding one for models of class II) was integrated
numerically in all the cases. Note that the outcome of
(\ref{eqdeltaH}) does not depend on the number of data points in
the interval.

\begin{figure*}[htbp]
    \centering
        \includegraphics[width=0.31\linewidth]{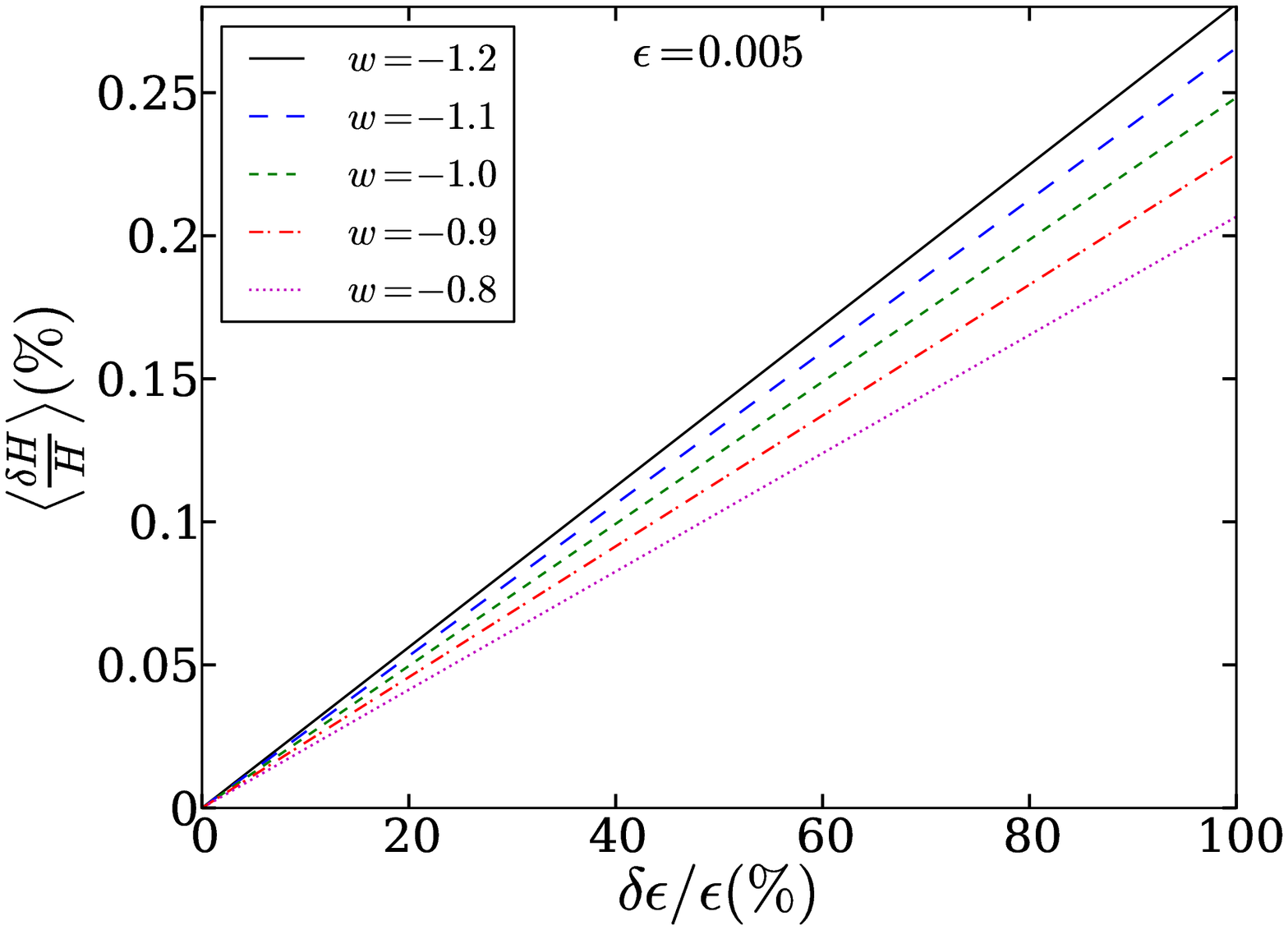}
        \includegraphics[width=0.31\linewidth]{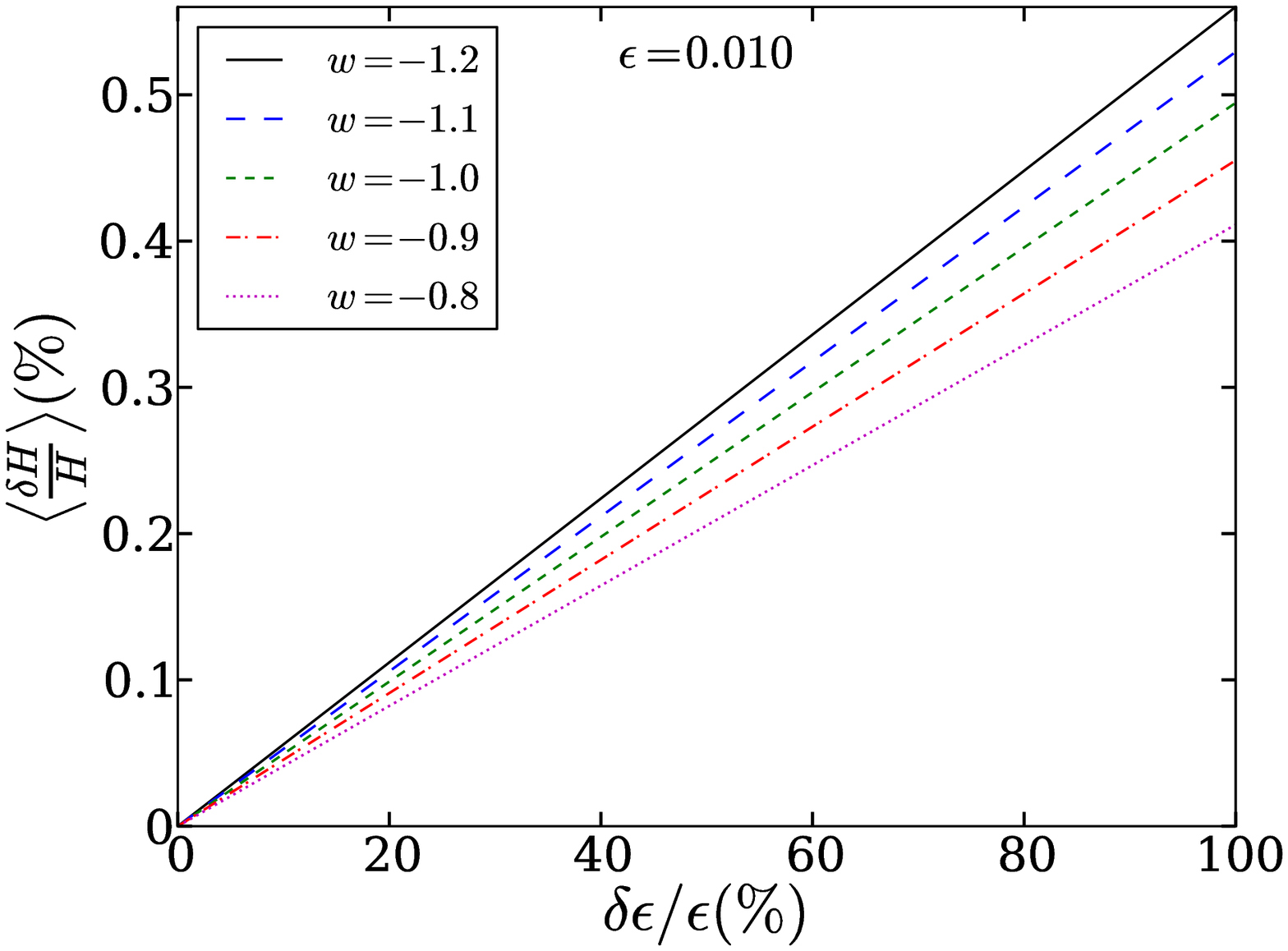}
        \includegraphics[width=0.31\linewidth]{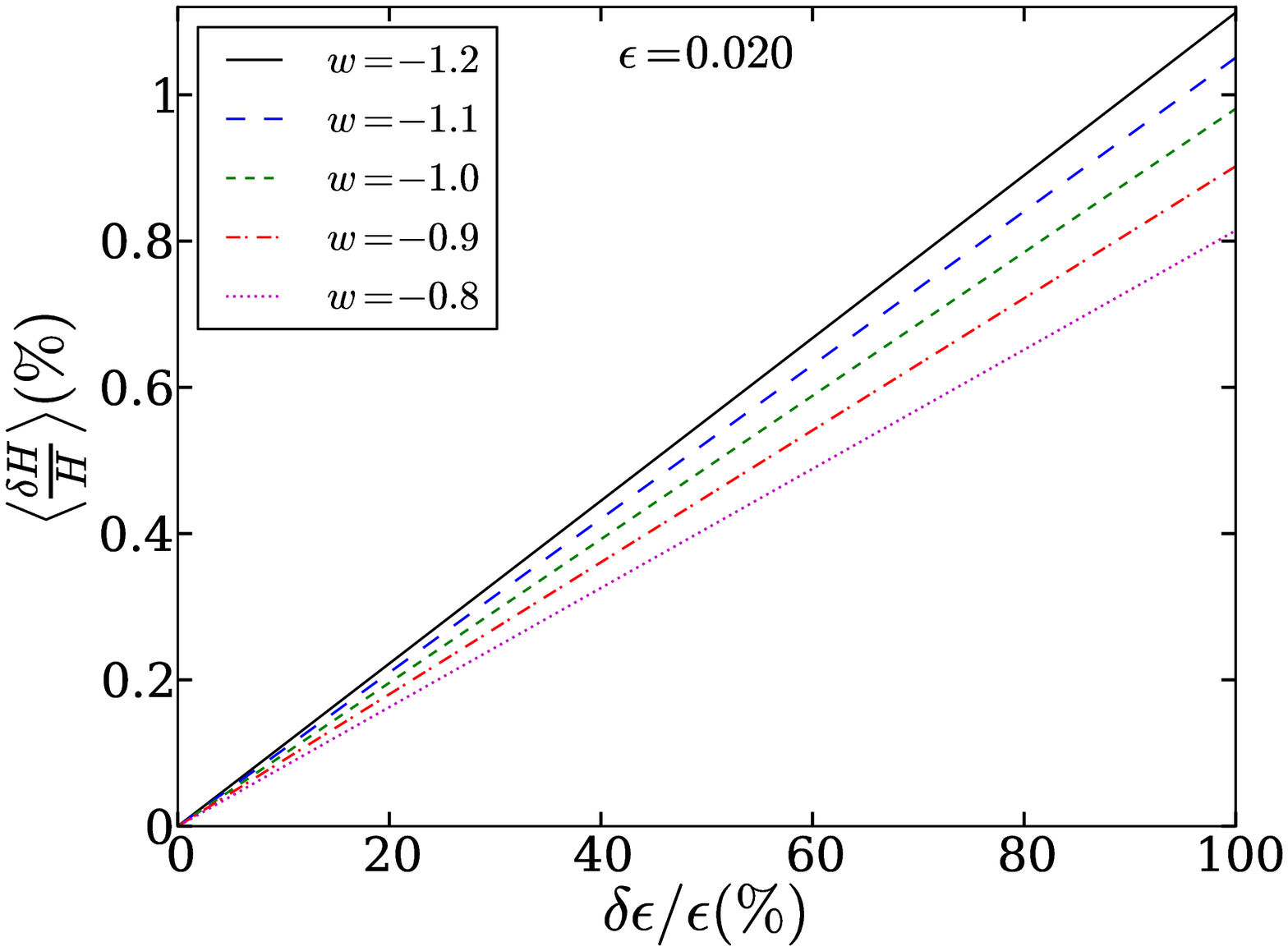}
    \caption{Estimates of $ \left< \frac{\delta H}{H} \right> $ for {\em Ansatz} $Q_{1a}$,
    Eq. (\ref{eq:Q1a}) for three different values of the parameter $\epsilon$.}
    \label{fig:mean_dHde_Q1a}
    \centering
        \includegraphics[width=0.31\linewidth]{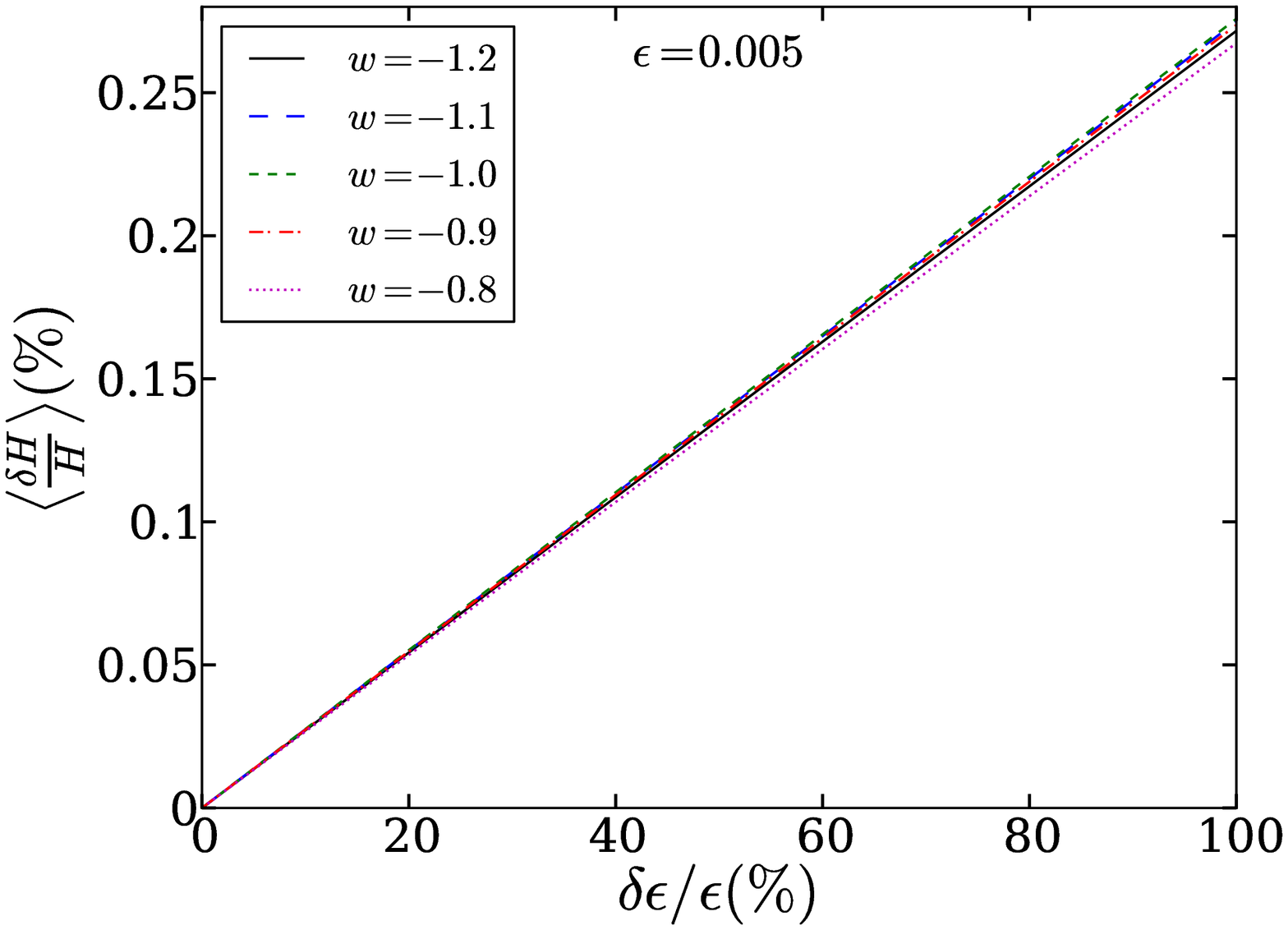}
        \includegraphics[width=0.31\linewidth]{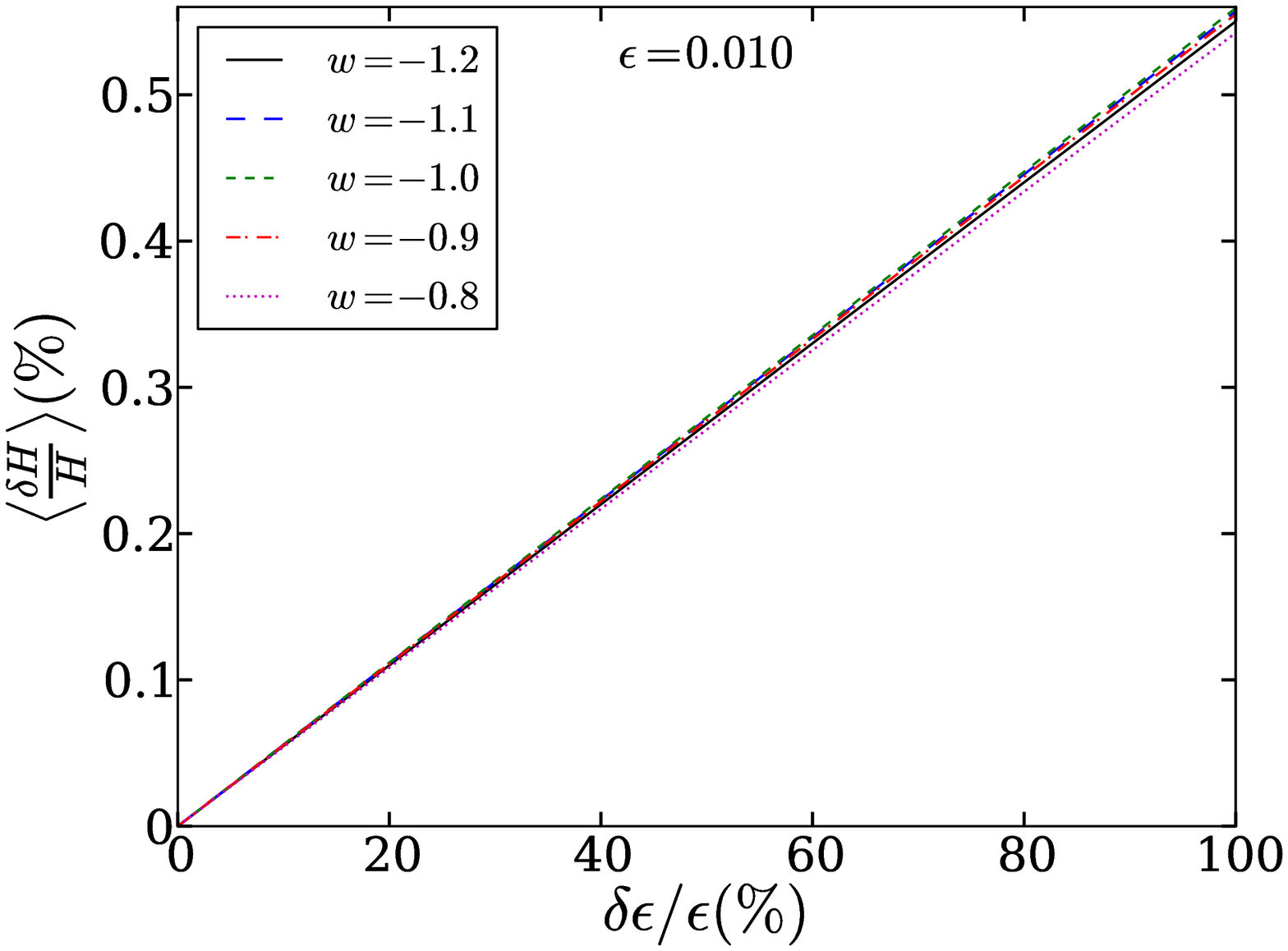}
        \includegraphics[width=0.31\linewidth]{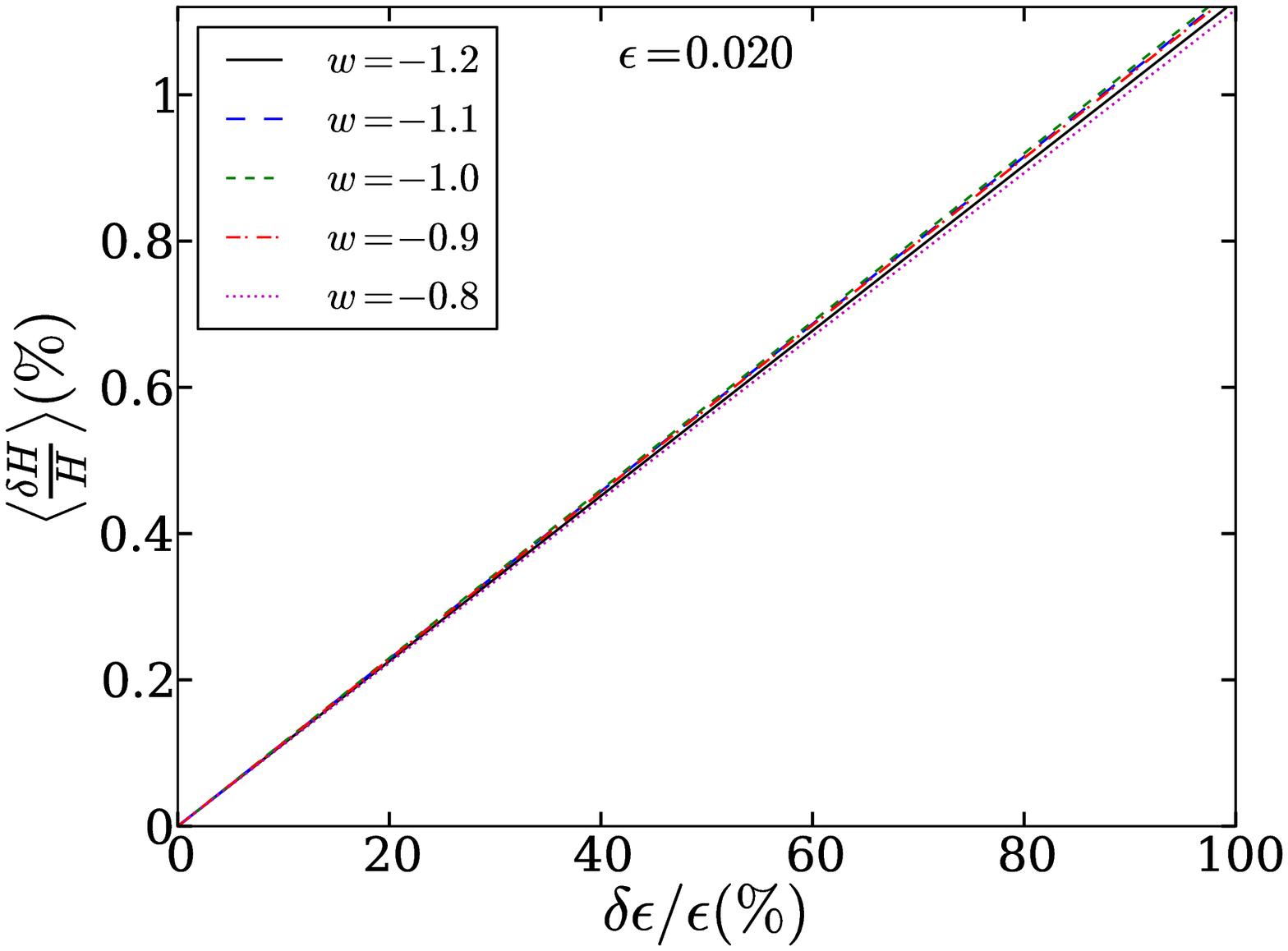}
    \caption{The same as Fig. \ref{fig:mean_dHde_Q1a}, but for model $Q_{1b}$, Eq. (\ref{eq:Q1b}).}
    \label{fig:mean_dHde_Q1b}
    \centering
        \includegraphics[width=0.31\linewidth]{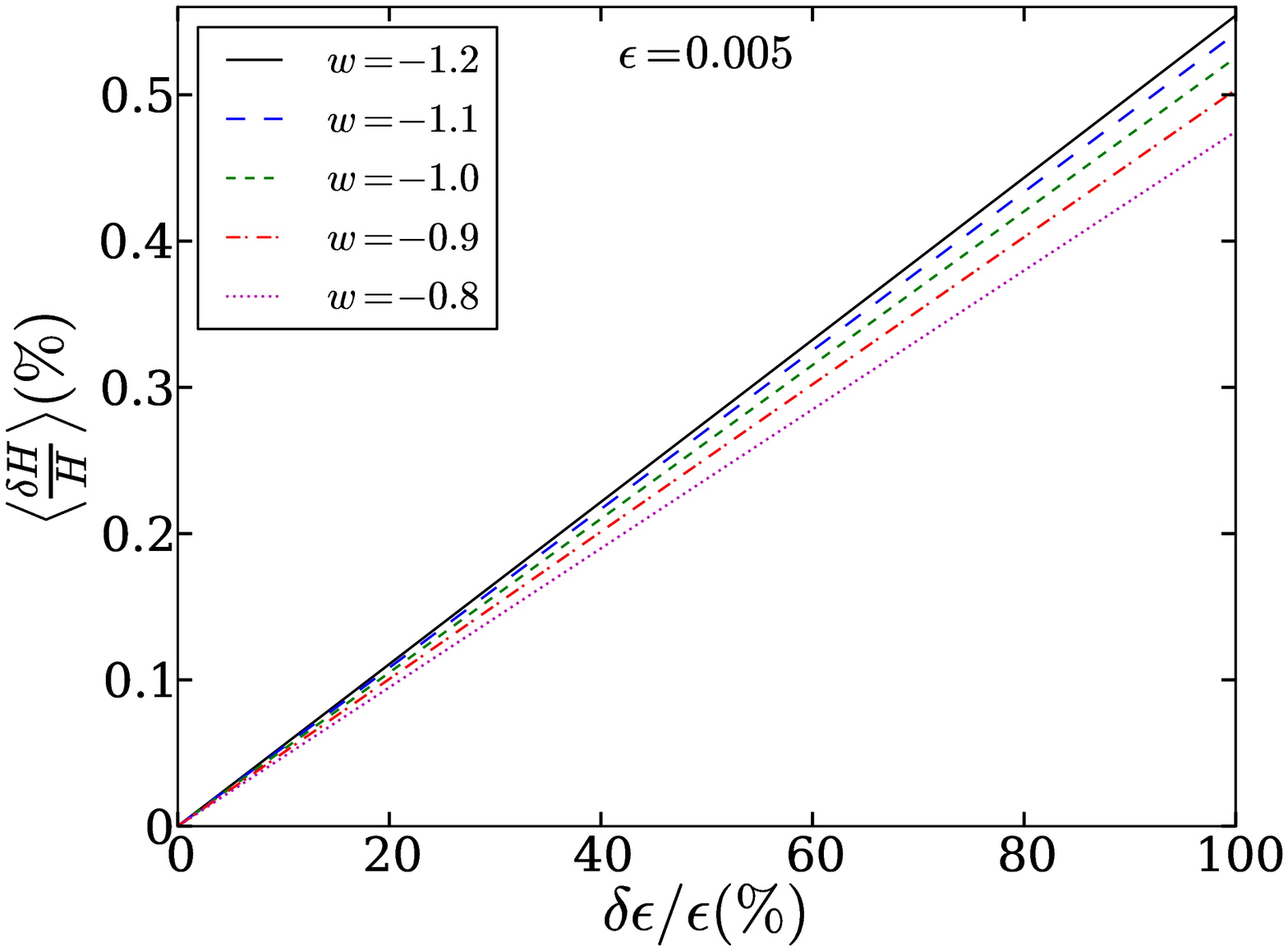}
        \includegraphics[width=0.31\linewidth]{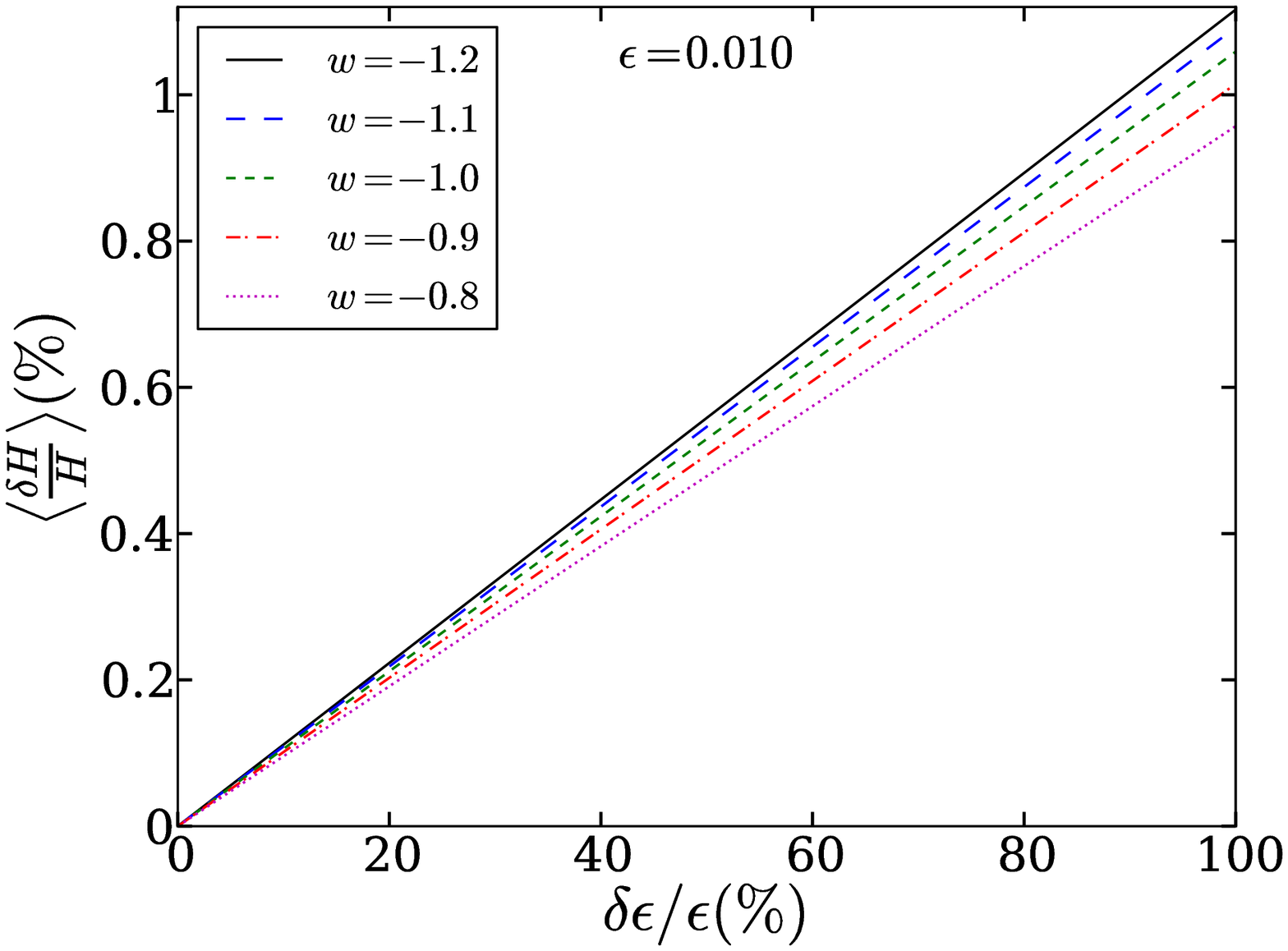}
        \includegraphics[width=0.31\linewidth]{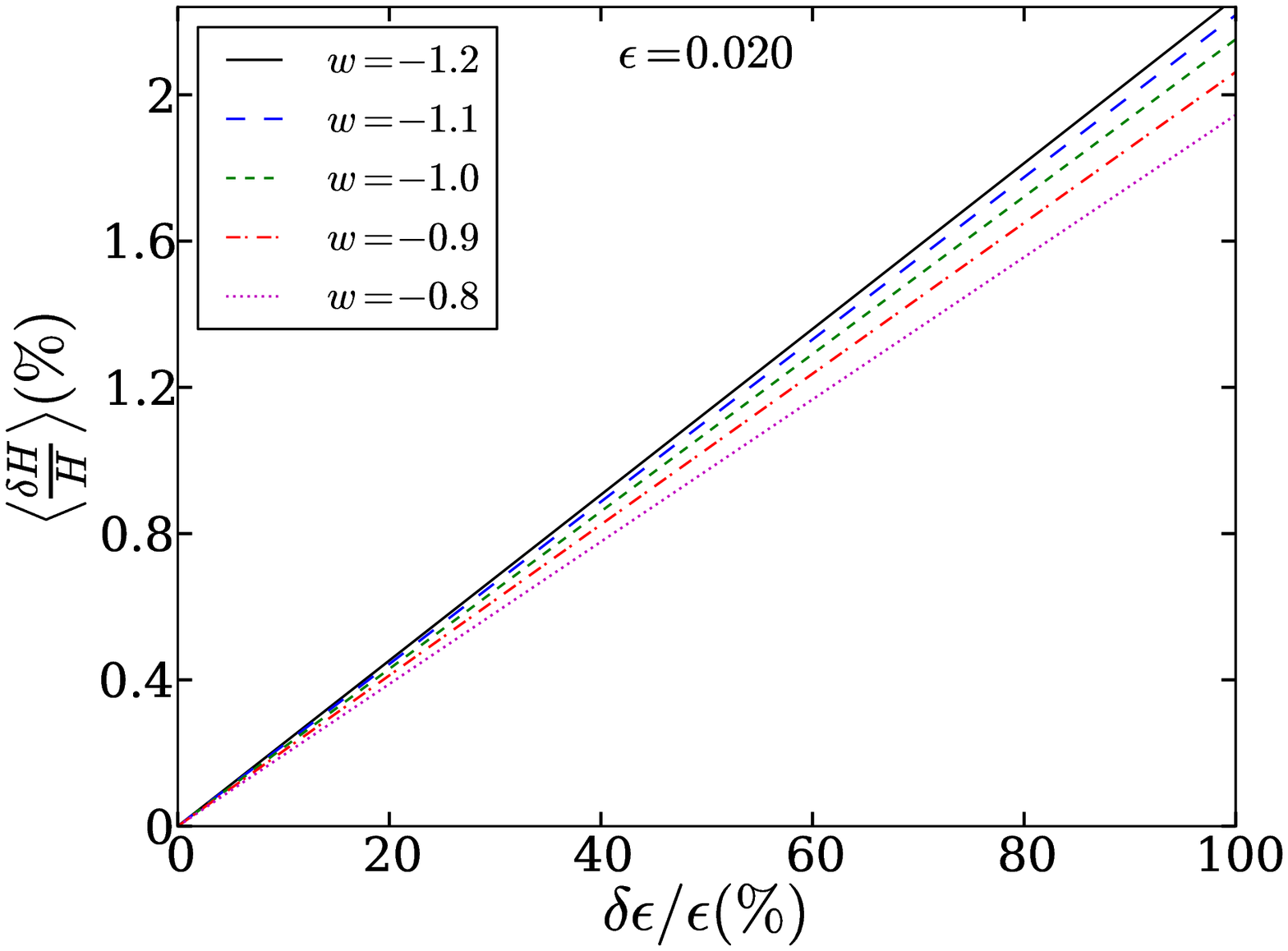}
    \caption{The same as Fig. \ref{fig:mean_dHde_Q1a}, but for model $Q_{1c}$, Eq. (\ref{eq:Q1c}).}
    \label{fig:mean_dHde_Q1c}
\end{figure*}

\begin{figure*}[htbp]
    \centering
        \includegraphics[width=0.31\linewidth]{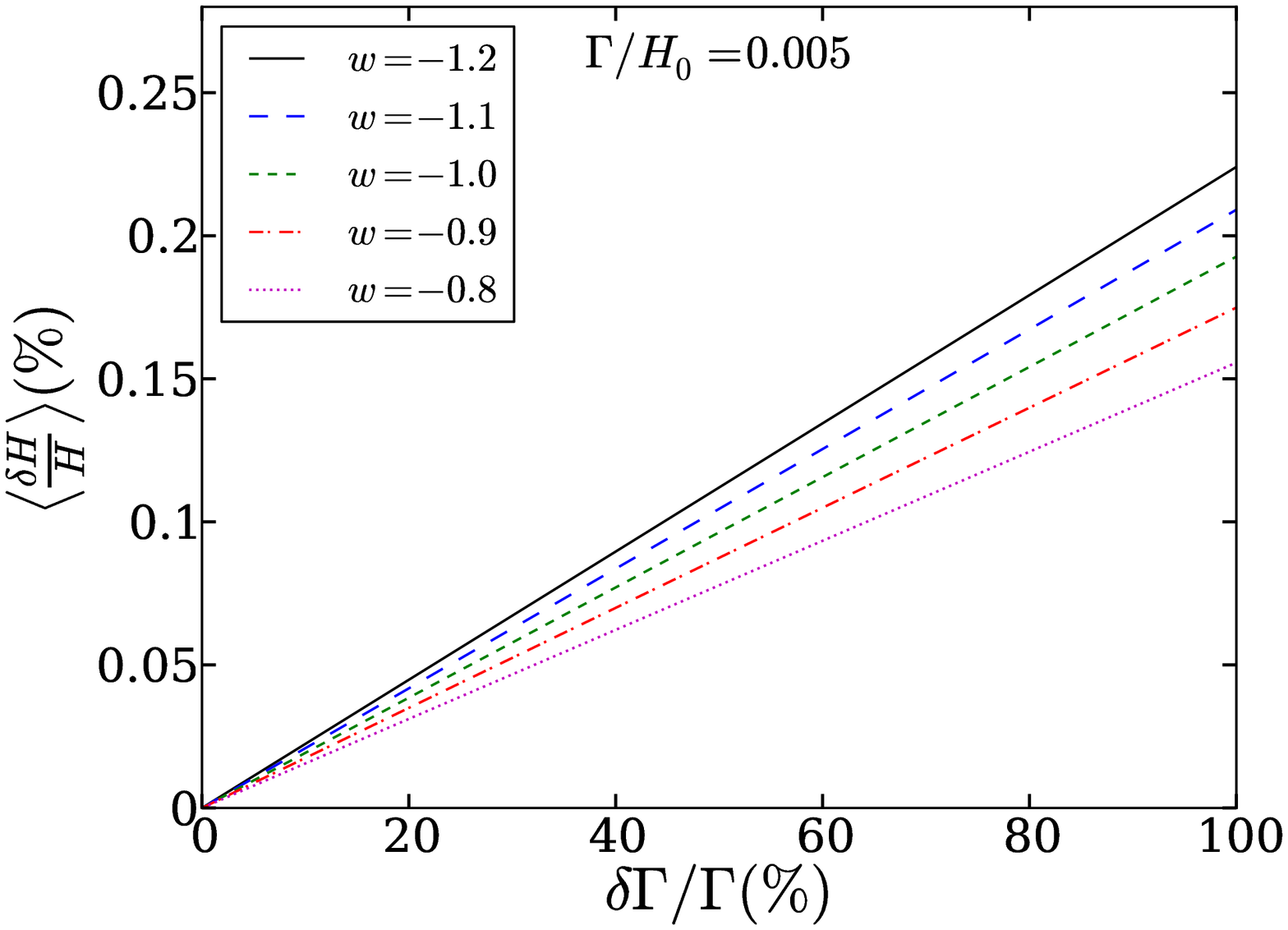}
        \includegraphics[width=0.31\linewidth]{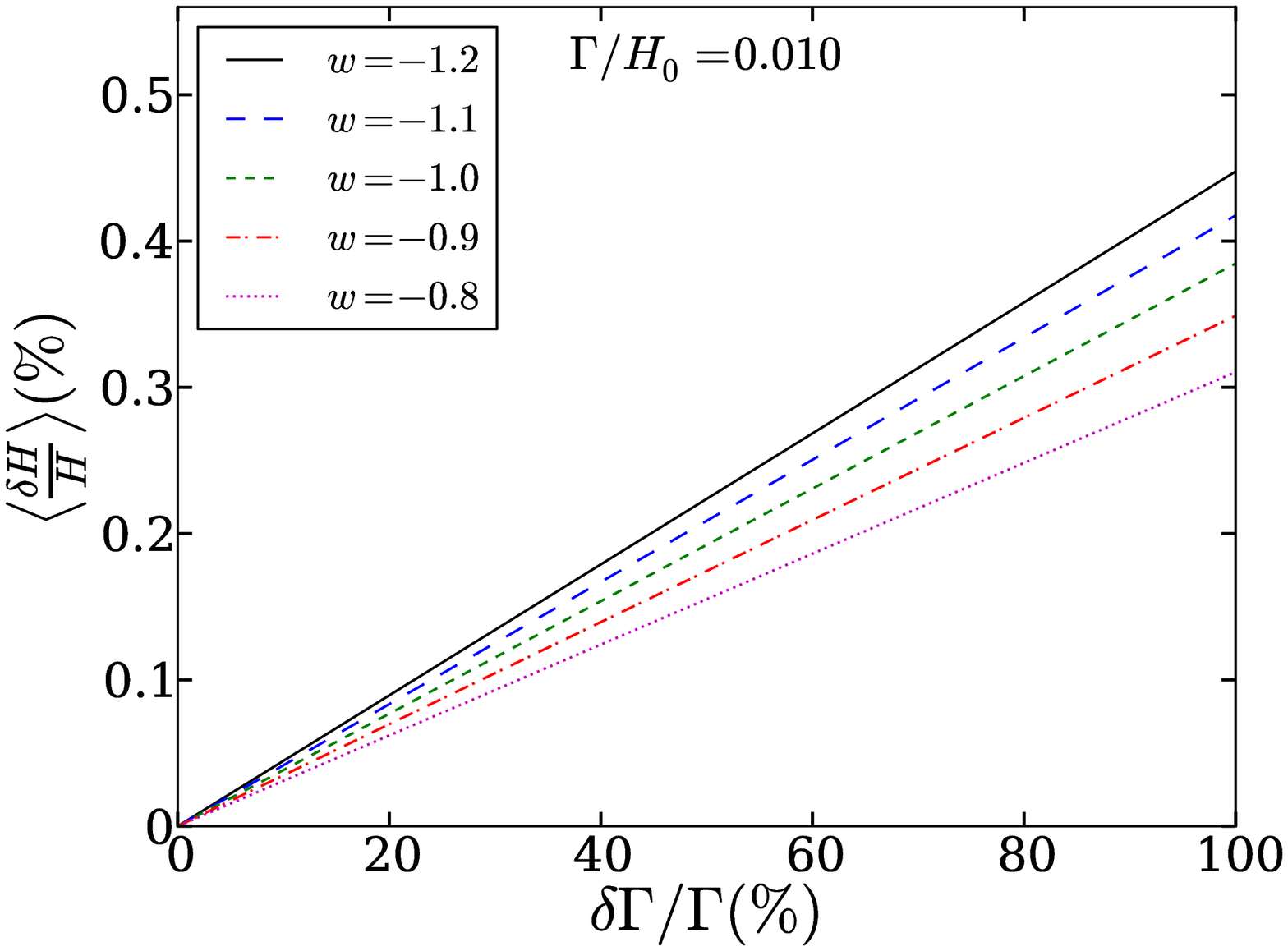}
        \includegraphics[width=0.31\linewidth]{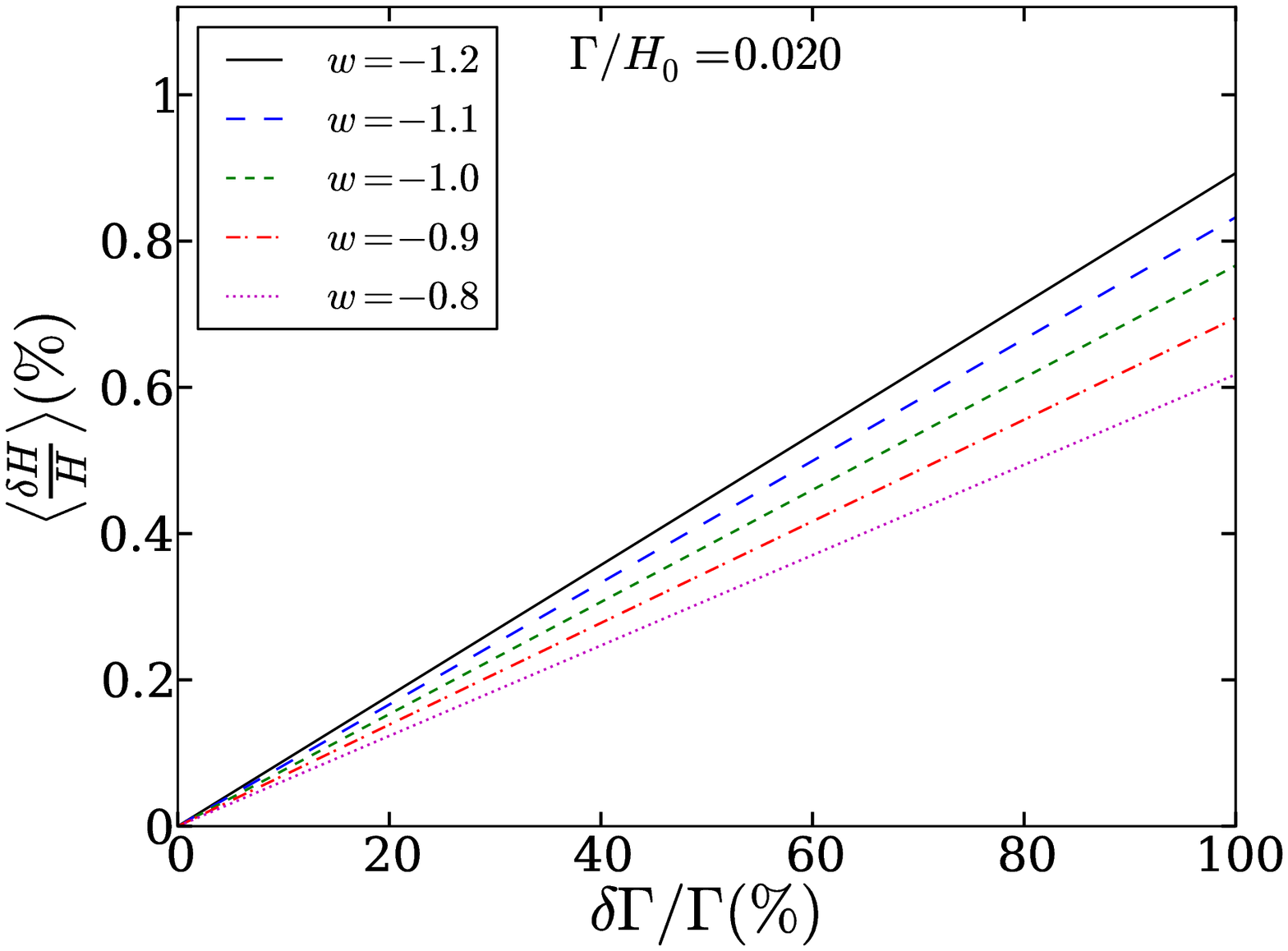}
    \caption{Estimates of $ \left< \frac{\delta H}{H} \right> $ for {\em Ansatz} $Q_{2a}$, Eq. (\ref{eq:Q2a})
    for three different values of the parameter $\Gamma / H_0$.}
    \label{fig:mean_dHde_Q2a}
    \centering
        \includegraphics[width=0.31\linewidth]{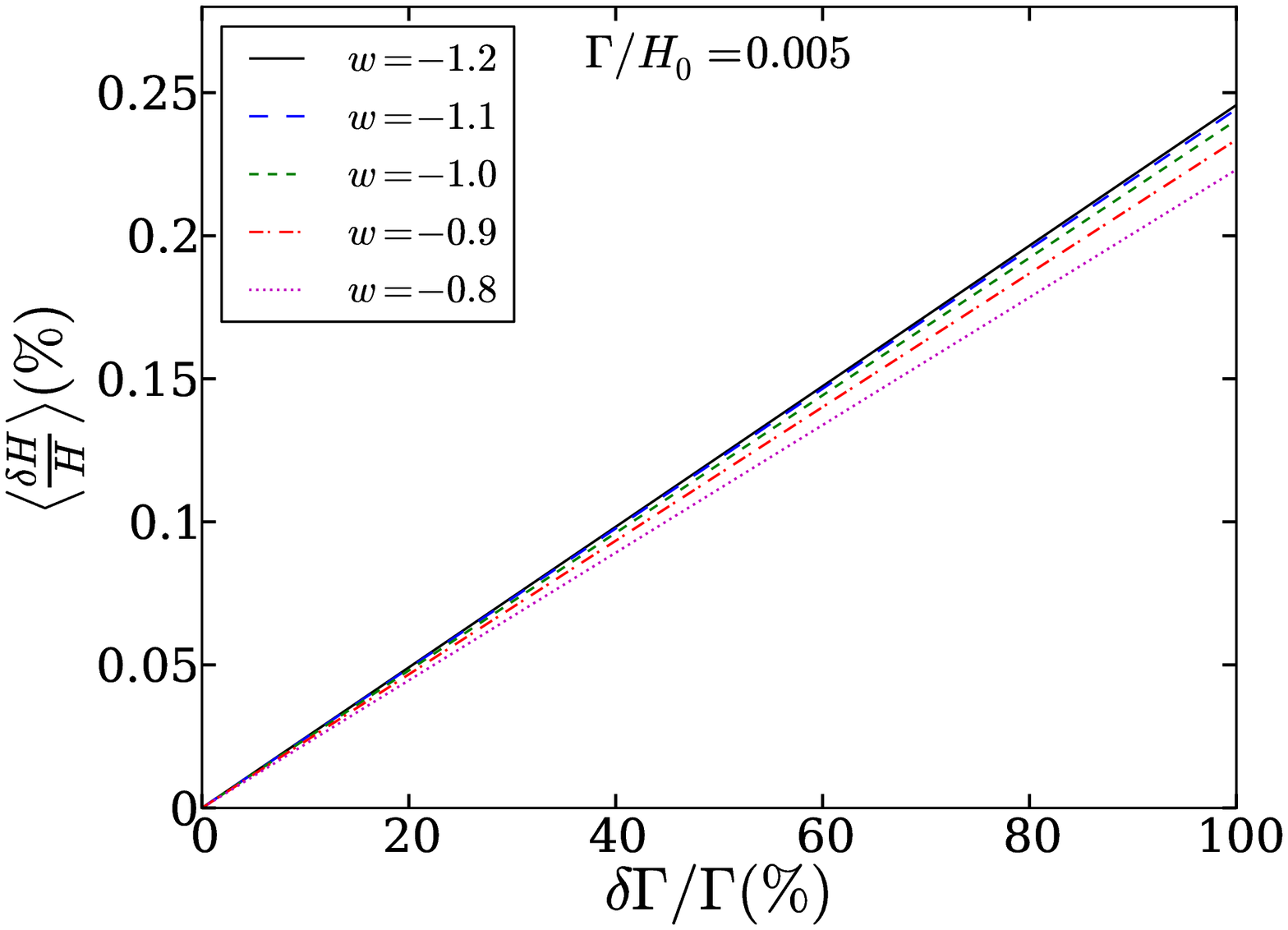}
        \includegraphics[width=0.31\linewidth]{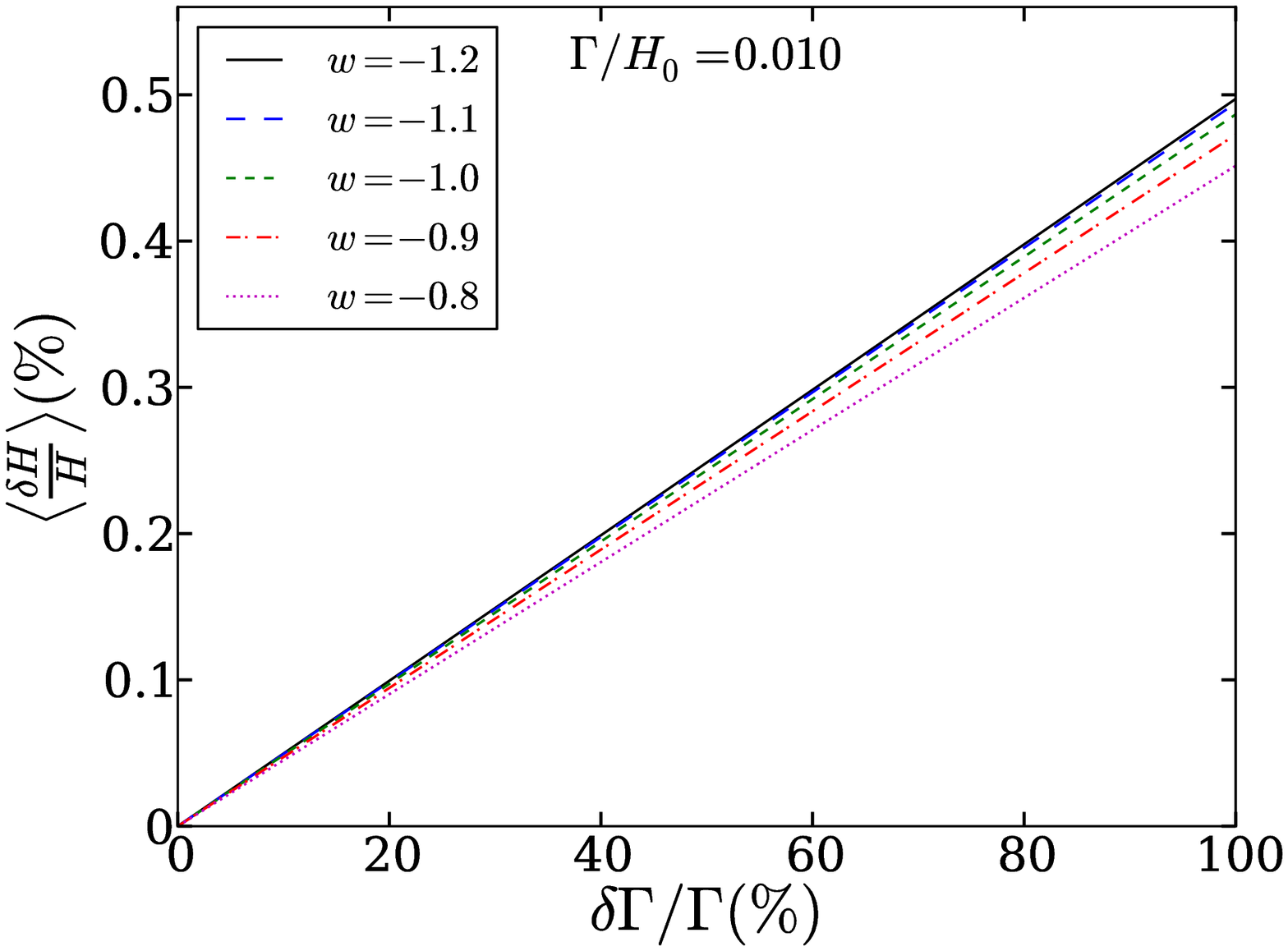}
        \includegraphics[width=0.31\linewidth]{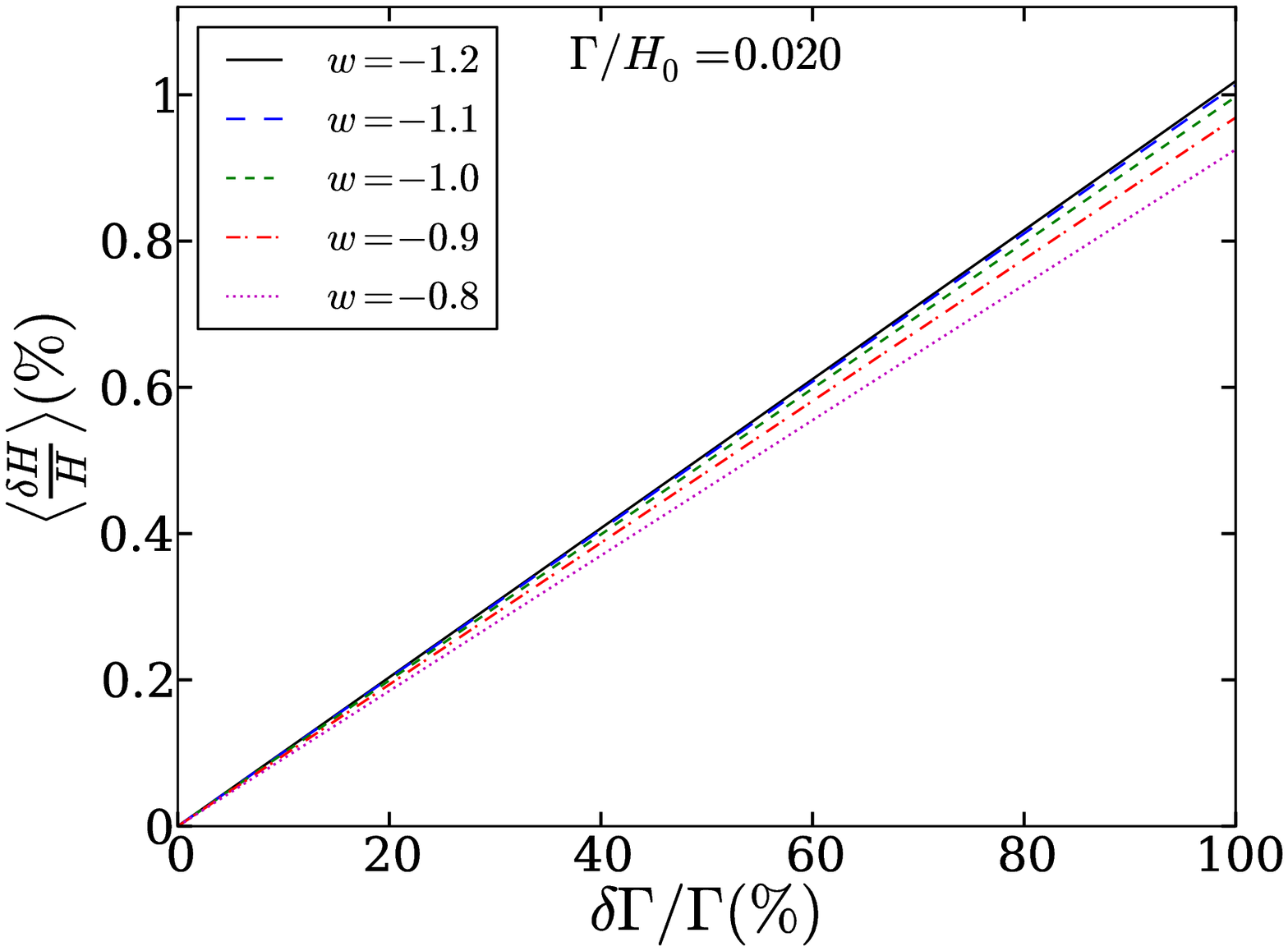}
    \caption{The same as Fig. \ref{fig:mean_dHde_Q2a}, but for ansatz $Q_{2b}$, Eq. (\ref{eq:Q2b}).}
    \label{fig:mean_dHde_Q2b}
    \centering
        \includegraphics[width=0.31\linewidth]{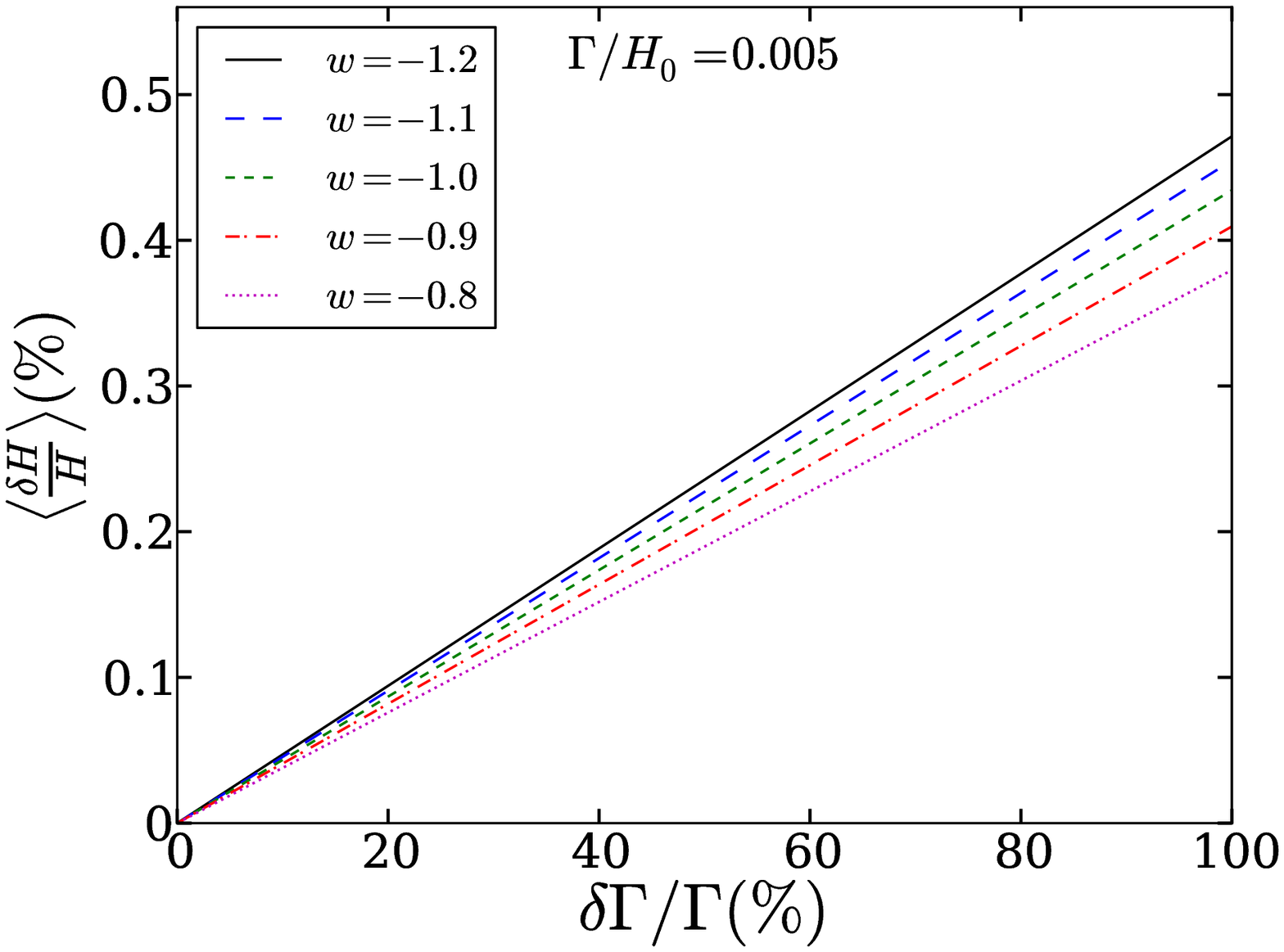}
        \includegraphics[width=0.31\linewidth]{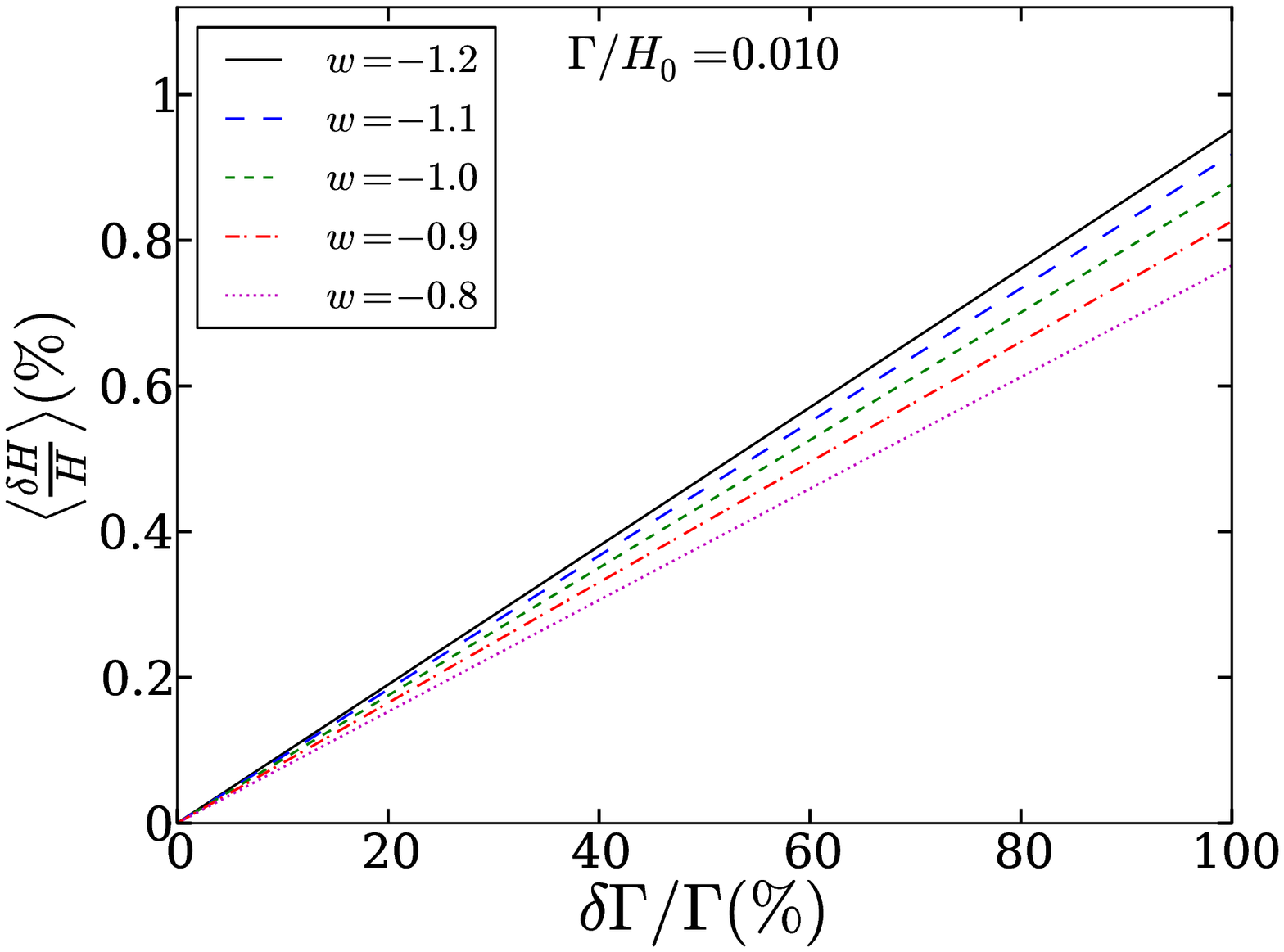}
        \includegraphics[width=0.31\linewidth]{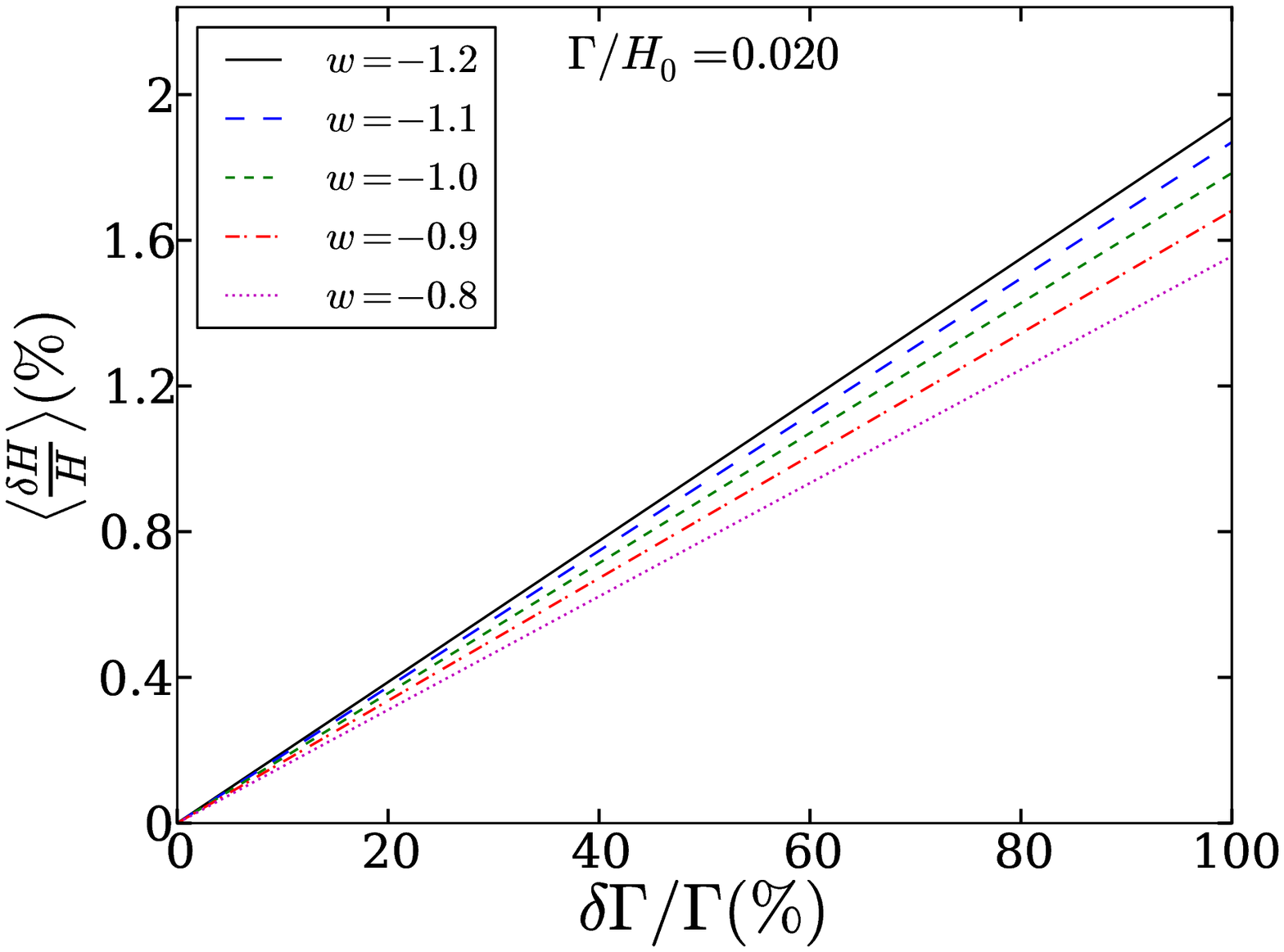}
    \caption{The same as Fig. \ref{fig:mean_dHde_Q2a}, but for {\em Ansatz} $Q_{2c}$, Eq. (\ref{eq:Q2c}).}
    \label{fig:mean_dHde_Q2c}
\end{figure*}

The results for models of type I are plotted in Figs.
\ref{fig:mean_dHde_Q1a},  \ref{fig:mean_dHde_Q1b} and
\ref{fig:mean_dHde_Q1c} for {\em Ans\"{a}tze} $Q_{1a}$, $Q_{1b}$
and $Q_{1c}$, respectively,  for $\epsilon = 0.005, 0.01, 0.02$
and five values of $w$. Likewise, the outcomes for models of type
II are plotted in Figs. \ref{fig:mean_dHde_Q2a},
\ref{fig:mean_dHde_Q2b}, and \ref{fig:mean_dHde_Q2c} for $Q_{2a}$,
$Q_{2b}$, and $Q_{2c}$, respectively, for $\Gamma/H_0$ and $w$
with the same values as for models of type I. Note that the
limiting case where it is possible to tell an interaction with
$1\sigma$ confidence level corresponds to $\delta \epsilon /
\epsilon = 100\%$ ($\delta \Gamma / \Gamma = 100\%$).

%%%%%%%%%%%%%%%%%%%%%%%%%%%%%%%%%%%%%%%%%%%%%%%%%%%%%%%%%%%%%%%%%%%%%%%%%%%%%%%%%%%%%%%%%%%%%%%%%%%%%%%
%%%%%%%%%%%%%%%%%%%%%%%%%%%%%%%%%%%%%%%%%%%%%%%%%%%%%%%%%%%%%%%%%%%%%%%%%%%%%%%%%%%%%%%%%%%%%%%%%%%%%%%
%%%%%%%%%%%%%%%%%%%%%%%%%%%%%%%%%%%%%%%%%%%%%%%%%%%%%%%%%%%%%%%%%%%%%%%%%%%%%%%%%%%%%%%%%%%%%%%%%%%%%%%
%%%%%%%%%%%%%%%%%%%%%%%%%%%%%%%%%%%%%%%%%%%%%%%%%%%%%%%%%%%%%%%%%%%%%%%%%%%%%%%%%%%%%%%%%%%%%%%%%%%%%%%
%%%%%%%%%%%%%%%%%%%%%%%%%%%%%%%%%%%%%%%%%%%%%%%%%%%%%%%%%%%%%%%%%%%%%%%%%%%%%%%%%%%%%%%%%%%%%%%%%%%%%%%
%%%%%%%%%%%%%%%%%%%%%%%%%%%%%%%%%%%%%%%%%%%%%%%%%%%%%%%%%%%%%%%%%%%%%%%%%%%%%%%%%%%%%%%%%%%%%%%%%%%%%%%
%%%%%%%%%%%%%%%%%%%%%%%%%%%%%%%%%%%%%%%%%%%%%%%%%%%%%%%%%%%%%%%%%%%%%%%%%%%%%%%%%%%%%%%%%%%%%%%%%%%%%%%
%%%%%%%%%%%%%%%%%%%%%%%%%%%%%%%%%%%%%%%%%%%%%%%%%%%%%%%%%%%%%%%%%%%%%%%%%%%%%%%%%%%%%%%%%%%%%%%%%%%%%%%

A feature common to Figs.
\ref{fig:mean_dHde_Q1a}-\ref{fig:mean_dHde_Q2c} is that the more
negative $w$ is, the easier it is to detect the interaction (the
higher the slope of the graph). An exception is, however, $Q_{1b}$
(Fig. \ref{fig:mean_dHde_Q1b}), where the case $w=-1.1$ is
slightly easier to detect than $w=-1.2$, even though the $w$
dependence is considerably weaker than in the other cases. It is
also apparent that for the same value of the interaction parameter
($\epsilon$ or $\Gamma /H_0$) it is always easier to distinguish
the interaction in models with $Q \propto (\rho_m + \rho_x)$
($Q_{1c}$ and $Q_{2c}$), which gives $\delta H/H \sim 2$ times
larger than for models where $Q \propto \rho_m$ or $Q \propto
\rho_x$.

He, Wang, and Abdalla  \cite{He2011} determined, for model
$Q_{1b}$ (with $w <-1$), that $\epsilon = 0.024^{+0.034}_{-0.027}$
(see Table \ref{table_epsilon}). According to Fig.
\ref{fig:mean_dHde_Q1b} (right panel), it would  required an
accuracy of about $1.2\%$ in $H(z)$ to measure $\epsilon$ with
similar accuracy. Other constraints obtained by the same authors
(see Table \ref{table_epsilon}) are considerably more stringent
than that for {\em Ansatz} $Q_{1b}$ ($w <-1$). Another example: In
order to measure $\epsilon$ for model $Q_{1a}$ with the same
accuracy as in \cite{Costa2010} ($\epsilon = 0.01 \pm 0.01$,
equivalent to $\delta \epsilon / \epsilon = 1$; see Table
\ref{table_epsilon}), from Fig. \ref{fig:mean_dHde_Q1a} (middle
panel), an accuracy of $0.5\%$ (for $w=-1$) would be required in
measuring $H(z)$.

The results for model II (Figs. \ref{fig:mean_dHde_Q2a}-
\ref{fig:mean_dHde_Q2c}) are very similar to those of model I
(Figs. \ref{fig:mean_dHde_Q1a}-\ref{fig:mean_dHde_Q1c}), though a
slightly better accuracy is needed. For instance, an interaction $
\Gamma/H_{0} = 0.024$ for {\em Ansatz} $Q_{2b}$ could be detected
if $ \left<\delta H/H \right> \simeq 1\%$ (right panel of Fig.
\ref{fig:mean_dHde_Q2b}), instead of $ 1.2\% $ in the analogous
case of {\em Ansatz} $Q_{1b}$ described above.

\begin{figure}[htbp]
    \centering
        \includegraphics[width=0.4\linewidth]{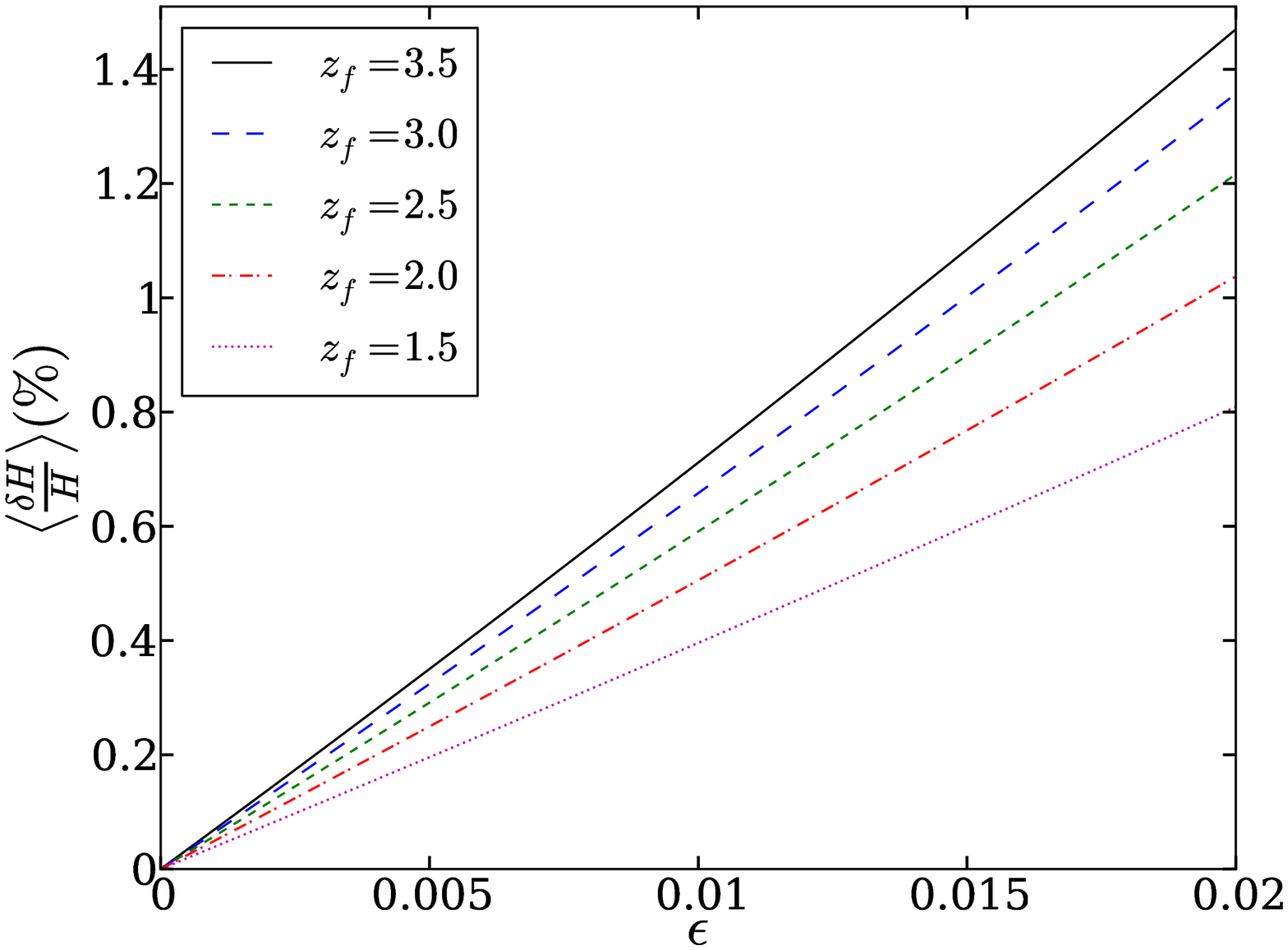}
    \caption{Dependence of $ \left< \frac{\delta H}{H} \right> $ with the redshift range of
    the data set for $\delta \epsilon = \epsilon$, and {\em Ansatz} $Q_{1b}$. In plotting the
    graphs we used $w = -1$, $z_{i} = 0.07$ and $\delta \epsilon / \epsilon = 100\%$.}
    \label{fig:mean_dHde_zf_Q1b}
\end{figure}

We next consider how the redshift interval of a data set affects
the accuracy required to detect an interaction. Figure
\ref{fig:mean_dHde_zf_Q1b} plots $ \left< \frac{\delta H}{H}
\right> $ in terms of $\epsilon$ for the case of {\em Ansatz}
$Q_{1b}$ for different values of the redshift range of the data.
It is seen that  the interaction (if it exists) is easier to
detect by observing H(z) at higher redshifts. Recently $H$ was
measured at $z = 2.3$ \cite{Busca2013}, extending the previous
upper value determined at $z=1.75$ \cite{Simon2005}. For instance,
an interaction such that $\epsilon = 0.02$ could be detected with
an $H(z)$ accuracy of $0.8\%$ for $z_f = 1.5$, whereas if $z_f$
were increased to $3.0$, an accuracy of $ \sim 1.3\%$ would
suffice to detect the interaction. This underlines the importance
of measuring $H$ at high redshifts.

%%%%%%%%%%%%%%%%%%%%%%%%%%%%%%%%%%%%%%%%%%%%%%%%%%%%%%%%%%%%%%%%%%%%%%%%%%%%%%%%%%%%%%%%%%%%%%
%%%%%%%%%%%%%%%%%%%%%%%%%%%%%%%%%%%%%%%%%%%%%%%%%%%%%%%%%%%%%%%%%%%%%%%%%%%%%%%%%%%%%%%%%%%%%%%

%%%%%%%%%%%%%%%%%%%%%%%%%%%%%%%%%%%%%%%%%%%%%%%%%%%%%%%%%%%%%%%%%%%%%%%%%%%%%%%%%%%%%%%%%%%%%%%
\subsection{Future measurements of H(z)}
The most recent $H(z)$ data \cite{Farooq2013} have an average
accuracy of only $\sim 12\%$. This shows how far we are from
distinguishing an interaction in the dark sector from the
noninteracting case using just $H(z)$ data. Next we briefly review
what should be expected from planned surveys.

First, a high accuracy is expected for $H_0$. The James Webb Space
Telescope, to be launched in 2018, will measure light curves of
Cepheid stars further than $50$ Mpc away, thus increasing the
number of supernovae calibrators, and may achieve a $1\%$
measurement of $H_0$ \cite{Gardner2010}.

Crawford \textit{et al.} estimated the observing time required for
the Southern African Large Telescope to measure $H(z)$ for
different accuracies \cite{Crawford2010}. By using the
differential ages method for luminous red galaxies with an
extended star formation history, they found that each redshift bin
measurement of $H(z)$ would require about  $180$ h to reaches
$\delta H/H = 3\%$. That means an accuracy of $3\%$ may be reached
at present, but it would imply a considerable amount of
observation time. More promising measures of $H(z)$, using baryon
acoustic oscillations, are to come with the Wide Field Infrared
Survey Telescope (WFIRST) \cite{WFISRT2013}, to be launched in
2020. It will allow one to determine $H(z)$ with an aggregate
precision of $0.72\%$ in the redshift interval $1 < z < 2$ and
$1.8\%$ at $2 < z < 3$. This combination of high accuracy and
redshift range will offer a great opportunity not just to probe a
likely interaction in the dark sector, but also to test a wide
variety of cosmological models.

%%%%%%%%%%%%%%%%%%%%%%%%%%%%%%%%%%%%%%%%%%%%%%%%%%%%%%%%%%%%%%%%%%%%%%%%%%%%%%%%
%%%%%%%%%%%%%%%%%%%%%%%%%%%%%%%%%%%%%%%%%%%%%%%%%%%%%%%%%%%%%%%%%%%%%%%%%%%%%%%%
%%%%%%%%%%%%%%%%%%%%%%%%%%%%%%%%%%%%%%%%%%%%%%%%%%%%%%%%%%%%%%%%%%%%%%%%%%%%%%%%
\section{Concluding remarks}
In this paper we considered two previously studied classes of
interacting DM-DE, Eqs. (\ref{eq:model1}) and (\ref{eq:model2}),
and used their one-parameter special cases, Eqs.
(\ref{eq:Q1a})-(\ref{eq:Q2c}), to study (i) their ability to
address the coincidence problem (Sec. III), and (ii) the
possibility of detecting an interaction based solely on $H(z)$
observations (Sec. IV). We first studied the effect of the
mentioned dark sector interaction forms in the evolution of the
ratio $\rho_m / \rho_x$ and  confirmed that a transfer of energy
from DM to DE (negative $\epsilon$ or $\Gamma$) only makes the
coincidence problem more severe. The coincidence problem gets more
alleviated with increasing $\epsilon$ (or $\Gamma$). The {\em
Ans\"{a}tze} that fare better with that issue when account is
taken also on the bounds on early dark energy \cite{xia-viel} are
$Q_{1b}$ for models of class I and $Q_{2b}$ and $Q_{2c}$ for class
II. Likewise, phantom models worsen the  problem. A curious fact
is that models of type I and II exhibit a like  behavior for
similar values  $\epsilon$ and $\Gamma/H_0$, despite the fact that
class I depends directly on $H(z)$ while class II  does not. We
also investigated the influence of the (constant) equation of
state parameter $w$ and showed that the coincidence problem gets
more alleviated when $w > -1$, though not significantly. On the
other hand, phantom cases require smaller values of $\Gamma/H_{0}$
to avoid negative densities in the futures (for {\em Ans\"{a}tze}
$Q_{2b}$ and $Q_{2c}$).

We do not claim that models that more alleviate the coincidence
problem are closer to the ``true'' model, but they certainly are
more appealing and therefore deserve special attention, from both
the theoretical and observational viewpoints. Nevertheless, we are
far from a complete picture.

We calculated how the errors in $H(z)$ propagate to the
interaction parameters and made several estimates of the
possibilities of measuring an interaction, depending on the {\em
Ansatz} employed (Sec. IV). Models with $Q \propto (\rho_m +
\rho_x)$ are detectable with less accuracy (about $2$ times worse)
than for the same value of the interaction parameter of models
with $Q \propto \rho_m$ or $Q \propto \rho_x$. Likewise, the
equation of state parameter of DE, $w$, may have a considerable
influence on the value of $ \left< \frac{\delta H}{H} \right> $.
In particular, for $w <-1$ the  interaction is easier to detect
(but, as said above, it worsens the coincidence problem and
suffers from others severe drawbacks). We found that if the WFIRST
experiment reach an average accuracy of $0.7\%$ up to $z = 3.0$,
it will be able to see an interaction with $1 \sigma$ confidence
level of $\epsilon \gtrsim 0.01$ (for $w=-1$) (modulo the
interaction exists). According to the constraints of for {\em
Ansatz} $Q_{1a}$ (Table \ref{table_epsilon}, Ref.
\cite{Costa2010}), this value is still allowed by observation.
Other model forecasts are discussed in Sec. IV.

We may conclude by saying that the $H(z)$ data should be
complemented with data from other sources (such as supernovae type
Ia, baryon acoustic oscillations, dynamical evolution of galaxy
clusters, integrated Sachs-Wolfe effect, growth factor, etc) to
determine with certainty whether DM and DE energy interact with
each other also nongravitationally.

Claims were made that in models of class I, with $w > -1$, density
perturbations blow up on super-Hubble scales \cite{valiviita}, no
matter how small the coupling parameter, $\epsilon$, might be. The
smaller $\epsilon$, the earlier the instability would set in. This
defies intuition because, in any case, in the $\epsilon
\rightarrow 0$ limit the perturbations should not diverge at all.
Recently, a detailed analysis of the energy-momentum transfer
between both dark components disproved the claims \cite{Sun2013}.

Admittedly, a clear limitation of this work is its restriction to
the $w$ constant. It would be interesting to adopt a
parametrization for this quantity, such as the one proposed by
Chevallier and Polarski \cite{chevallier} and Linder
\cite{linder}. However, it would introduce a further parameter
which would much involve the calculations. In any case, this may
well be the subject of a future research.

%%%%%%%%%%%%%%%%%%%%%%%%%%%%%%%%%%%%%%%%%%%%%%%%%%%%%%%%%%%%%%%%%%%%%%%%%%%%%%%%%%%%
%%%%%%%%%%%%%%%%%%%%%%%%%%%%%%%%%%%%%%%%%%%%%%%%%%%%%%%%%%%%%%%%%%%%%%%%%%%%%%%%%%%%
%%%%%%%%%%%%%%%%%%%%%%%%%%%%%%%%%%%%%%%%%%%%%%%%%%%%%%%%%%%%%%%%%%%%%%%%%%%%%%%%%%%%
%%%%%%%%%%%%%%%%%%%%%%%%%%%%%%%%%%%%%%%%%%%%%%%%%%%%%%%%%%%%%%%%%%%%%%%%%%%%%%%%%%%%
\begin{acknowledgments}
P.C.F. acknowledges Jailson Alcaniz for useful discussions at an
early stage of this paper; he also thanks the Department of
Physics of the ``Universidad Aut\'{o}noma de Barcelona", where
this work was done, for warm hospitality. P.C.F. acknowledges
financial support from  CAPES  Scholarship No. Bex 18138/12-8 and
CNPq. This research was partially supported by the ``Ministerio de
Econom\'{\i}a y Competitividad, Direcci\'{o}n General de
Investigaci\'{o}n Cient\'{\i}fica y T\'{e}cnica", Grant No.
FIS2012-32099. J.C.C. acknowledges support from the CNPq (Brazil).
\end{acknowledgments}
%%%%%%%%%%%%%%%%%%%%%%%%%%%%%%%%%%%%%%%%%%%%%%%%%%%%%%%%%%%%%%%%%%%%%%%%%%%%%%%%%%%%
%%%%%%%%%%%%%%%%%%%%%%%%%%%%%%%%%%%%%%%%%%%%%%%%%%%%%%%%%%%%%%%%%%%%%%%%%%%%%%%%%%%%
%%%%%%%%%%%%%%%%%%%%%%%%%%%%%%%%%%%%%%%%%%%%%%%%%%%%%%%%%%%%%%%%%%%%%%%%%%%%%%%%%%%%
%%%%%%%%%%%%%%%%%%%%%%%%%%%%%%%%%%%%%%%%%%%%%%%%%%%%%%%%%%%%%%%%%%%%%%%%%%%%%%%%%%%%

\end{document}